\documentclass[12pt,draftclsnofoot,onecolumn]{IEEEtran}

\usepackage{graphicx}
\usepackage{subfigure}
\usepackage{amsmath,amsthm,amsfonts}
\usepackage{cite}
\usepackage{bm}
\usepackage{url}
\usepackage{array}

\allowdisplaybreaks[4]
\theoremstyle{remark}
\newtheorem*{rem}{Remark}

\begin{document}

\title{Asynchronous Physical-layer Network Coding: Symbol Misalignment Estimation and Its Effect on Decoding}
\author{Yulin~Shao,
        Soung~Chang~Liew,~\IEEEmembership{Fellow,~IEEE,}
        and~Lu~Lu,~\IEEEmembership{Member,~IEEE}
\thanks{Y. Shao and S. C. Liew are with the Department of Information Engineering, The Chinese University of Hong Kong, Shatin, New Territories, Hong Kong. Email:~\{sy016,~soung\}@ie.cuhk.edu.hk. L. Lu is with the Institute of Network Coding, The Chinese University of Hong Kong, Shatin, New Territories, Hong Kong. Email:~lulu@ie.cuhk.edu.hk.}
}
\maketitle
\vspace{-2em}
\begin{abstract}
In practical asynchronous physical-layer network coding (APNC) systems, the symbols from multiple transmitters to a common receiver may be misaligned.
The knowledge of the amount of symbol misalignment, hence its estimation, is important to PNC decoding.
This paper addresses the problem of symbol-misalignment estimation and the problem of optimal PNC decoding given the misalignment estimate, assuming the APNC system uses the root-raised-cosine pulse to carry signals (RRC-APNC).
For symbol-misalignment estimation, we put forth an optimal estimator that is substantially more accurate than prior schemes.
Our estimator makes use of double baud-rate samples, and hence is information-lossless since the double baud-rate samples capture all the information embedded in the continuous-time signal shaped by the assumed RRC pulse.
Extensive simulations show that the mean-square-error (MSE) gains of our double baud-rate estimator over the baud-rate estimator are up to 8 dB.
For PNC decoding, we devise optimal APNC decoders under inaccurate symbol-misalignment estimates.
In particular, we prove and present a new whitening transformation to whiten the noise of the double baud-rate samples. Given the whitened samples, we construct a factor graph for a sum-product algorithm (SPA) to optimally decode the network-coded message.
We further investigate the decoding performance of various estimation-and-decoding solutions for RRC-APNC.
Simulation results indicate that:
(i) A large roll-off factor of the RRC pulse is preferable in order to reduce decoder complexity while maintaining good performance.
(ii) With a large roll-off factor, double baud-rate decoding is indispensable. Compared with a baud-rate sampling system (both estimation and decoding are based on baud-rate samples), a double baud-rate sampling system (both estimation and decoding are based on double baud-rate samples) yields $4.5$ dB gains on symbol error rate (SER) performance in an AWGN channel, and $2$ dB gains on packet error rate (PER) performance in a flat Rayleigh fading channel.
(iii) Under a large roll-off factor, our double baud-rate estimator yields considerable improvements on decoding performance over the baud-rate estimator assuming the double baud-rate decoder is used in both cases. In an AWGN channel, the SER performance gains are up to $2$ dB. In a flat Rayleigh fading channel, the PER performance gains are up to $1$ dB.
\end{abstract}


\section{Introduction}
Physical-layer Network Coding (PNC) improves throughput in wireless networks by turning interference into useful network-coded messages \cite{PNC1,PNC2}. When operated with PNC, the information exchange of a two-way relay network (TWRN) is divided into two phases: the uplink phase and the downlink phase. In the uplink phase, two end nodes transmit their messages to a relay simultaneously, and the relay tries to decode a network-coded message based on the overlapped signals received from the two end nodes. In the downlink phase, the relay broadcasts the network-coded message to both end nodes, and each end node extracts the other node's message using its pre-stored self-message and the received network-coded message \cite{PNCbook}.

PNC with asynchrony is termed Asynchronous PNC (APNC) \cite{luAPNC}. APNC does away with tight synchronization of the transmission times of the two end nodes. As a result, the transmitted signals from the two end nodes may arrive at the relay with a relative time offset (i.e., symbol misalignment). Recently, APNC that uses the root-raised-cosine (RRC) pulse has attracted much attention. Compared with APNC using the rectangular pulse \cite{PNCbook,luAPNC,BhatTWC}, APNC with the RRC pulse (RRC-APNC) is more spectrally efficient.

Research efforts on RRC-APNC have focused on two issues critical to its inner workings, namely, accurate symbol-misalignment estimation \cite{QingThesis,DangEstTau,QingEstTau} and optimal PNC decoding under Inter-Symbol Interference (ISI) \cite{DongnanBCJR,millerViterbi}. This paper makes advances on the solutions to these two problems, as elaborated below.

{\bf Symbol-misalignment estimation} -- Estimating the symbol misalignment is crucial in APNC because accurate estimate is required for good PNC decoding performance. Symbol misalignment estimators for RRC-APNC based on oversampling were proposed in \cite{QingThesis} and \cite{DangEstTau}. These estimators are highly complex owing to the oversampling. In \cite{DangEstTau}, simulations showed that as much as $8$-times oversampling is needed for accurate estimation. Our earlier work \cite{QingEstTau} presented a maximum likelihood (ML) estimator based on baud-rate sampling (i.e., no oversampling). This baud-rate estimator was shown to be able to achieve estimation accuracy comparable to that of the $8$-times oversampling estimators in \cite{QingThesis} and \cite{DangEstTau}.

An issue left open by \cite{QingEstTau} is that baud-rate sampling may not capture all the symbol misalignment information contained in the received signal when the RRC pulse is used. According to the Nyquist-Shannon sampling theorem, the Nyquist rate is the lower bound on the sampling rate that can guarantee no information loss when converting the continuous-time signal into the discrete-time signal (i.e., mathematically, the discrete-time signal can be converted back to the continuous-time signal exactly) \cite{Proakisbook}. For the signals shaped by the raised-cosine (RC) pulse or the RRC pulse, the Nyquist rate is between the baud-rate and the double baud-rate (i.e., it is defined by a roll-off factor $\beta$ within the range of $0\leq\beta\leq 1$). This observation suggests that (i) the baud-rate estimator in \cite{QingEstTau} may still be suboptimal and further improvement is possible with double baud-rate sampling for RRC-APNC; (ii) oversampling beyond the double-baud rate is not necessary because all signal information has already been captured with double baud-rate sampling.

With the above motivations, this paper puts forth an ML-optimal estimator that estimates symbol misalignment based on double baud-rate samples of the preamble. Simulation results show that this double baud-rate estimator substantially improves the symbol-misalignment estimation accuracy over the prior baud-rate estimator \cite{QingEstTau}. The mean-square-error (MSE) performance gains of the double baud-rate estimator over the baud-rate estimator are up to $8$ dB under both AWGN and Rayleigh fading channels.

{\bf Optimal decoders under ISI} -- In APNC, following the symbol-misalignment estimator, a PNC decoder will then make use of the estimated symbol misalignment to perform PNC decoding.
For a conventional point-to-point system using the RRC pulse, the optimal decoder is a symbol-by-symbol decoder since we could always sample at specific positions to eliminate the ISI \cite{Proakisbook}. However, for RRC-APNC, when the symbol boundaries of the two end nodes are not aligned, ISI is inevitable because there are no ISI-free sampling positions for both signals. Decoders designed for the ISI channel in RRC-APNC can be applied to harness useful information contained in the ISI. Two examples are the BCJR decoder in \cite{DongnanBCJR} and the MLSE decoder in \cite{millerViterbi}.
In these papers, perfect knowledge of the symbol misalignment is assumed as prior information to the receiver, and hence the decoders are designed under an ideal setup, i.e., all the samples are collected at the exact symbol boundaries of one of the two end nodes. However, in practice, the symbol misalignment needs to be estimated and the exact symbol boundaries may not be known perfectly.

To fill this gap, our paper here investigates the design of optimal RRC-APNC decoders in the presence of inaccurate symbol misalignment estimates. We construct a factor graph with a tree structure, from which a sum-product algorithm (SPA) can be devised to optimally decode the network-coded messages. In particular, for the APNC decoder using double baud-rate samples, we prove and present a new whitening transformation to whiten the colored noise. This whitening transformation involves only matrix decomposition and multiplication, hence is linear and computationally efficient.

{\bf Integration of symbol-misalignment estimation and PNC decoding} -- Prior work on APNC either focused on the issue of symbol-misalignment estimation or the issue of PNC decoding. The overall APNC system that takes into account the effect of symbol-misalignment estimation on PNC decoding was not studied. To fill this gap, this paper studies various integrated estimation-and-decoding solutions for RRC-APNC.
Specifically, there are four possible solutions at the receiver: baud-rate estimation and baud-rate decoding (solution I); double baud-rate estimation and baud-rate decoding (solution II); baud-rate estimation and double baud-rate decoding (solution III); double baud-rate estimation and double baud-rate decoding (solution IV).
Our comprehensive simulations on these estimation-and-decoding solutions indicate that:
(i) Practical RRC-APNC demands a large roll-off factor to enable feasible decoding when complexity is taken into account.
(ii) Under a large roll-off factor, double baud-rate decoding is indispensable for RRC-APNC. When $\beta=1$, solution IV yields $4.5$ dB symbol error rate (SER) gains over solution I in an AWGN channel, and $2.5$ dB packet error rate (PER) gains over solution I in a flat Rayleigh fading channel.
(iii) Under a large roll-off factor, solution IV yields considerable improvements on decoding performance over solution III. When $\beta=1$, the SER performance gains are up to $2$ dB in an AWGN channel, the PER performance gains are up to $1$ dB in a flat Rayleigh fading channel.

The remainder of this paper is organized in the following manner. Section II presents the integrated estimation-decoding framework for RRC-APNC. Section III details the baud-rate and double baud-rate symbol-misalignment estimators. Section IV devises the optimal RRC-APNC decoders under both baud-rate and double baud-rate sampling. Numerical results are also presented in Section III and IV. Section V concludes this paper. Throughout the paper, lowercase bold letters represent vectors while uppercase bold letters represent matrices.

\section{System Model}
\begin{figure}[h]
  \centering
  \includegraphics[width=0.9\columnwidth]{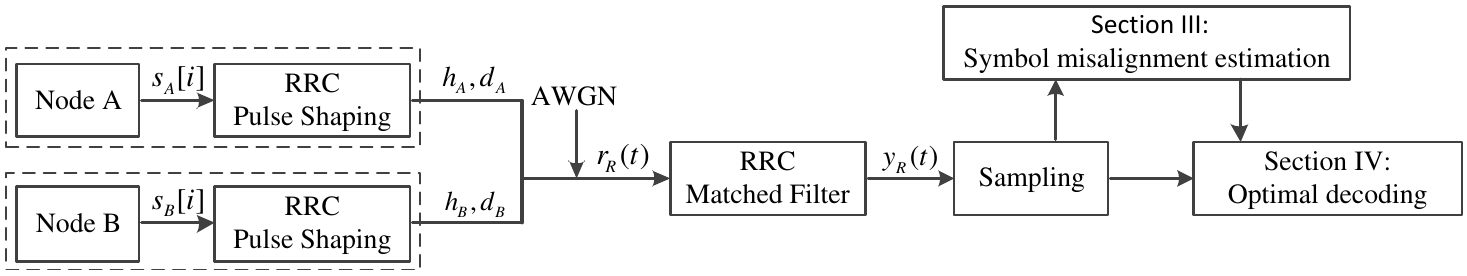}\\
  \caption{System model for the uplink of a TWRN operated with PNC where the symbols of nodes A and B arrive at the relay R with a relative time offset (i.e., symbol misalignment). We assume nodes A and B use practical RRC pulse shapes.}
\label{Fig1}
\end{figure}

We study a TWRN operated with PNC, where two end nodes A and B exchange information with the help of a relay R. Time domain RRC pulses are adopted by both nodes A and B. The system model is shown in Fig.\ref{Fig1}. In the PNC uplink, nodes A and B transmit signals to the relay R simultaneously, and there is a symbol misalignment between the signals of nodes A and B at relay R due to the imperfect synchronization of the arrival times of their symbols (i.e., the boundaries of the symbols from A and B are not aligned)~\cite{luAPNC}.

We denote by $\bm{s_A}$ and $\bm{s_B}$ the symbols to be transmitted at nodes A and B. After going through the RRC pulse-shaping filter, the continuous-time baseband signal at each node can be expressed as
\begin{eqnarray}
\label{EquA1}
x_A(t) \!\!&=&\!\! \sum_i s_A[i]p'(t-iT),\\
\label{EquA2}
x_B(t) \!\!&=&\!\! \sum_i s_B[i]p'(t-iT),
\end{eqnarray}
where $T$ is the symbol duration, and $p'(t)$ is an RRC pulse with a roll-off factor $\beta:0\leq\beta\leq 1$.

At relay R, the received overlapping baseband continuous-time signal is given by
\begin{eqnarray}\label{EquA3}
r_R(t) \!\!&=&\!\!h_Ax_A(t-d_A)+h_Bx_B(t-d_B)+n_R(t)\nonumber\\
&=&\!\!h_A\sum_i s_A[i]p'(t-iT-d_A)+h_B\sum_i s_B[i]p'(t-iT-d_B)+ n_R(t),
\end{eqnarray}
where $h_A$ and $h_B$ are the channel gains (in this paper, we assume slow-fading channel, and hence $h_A$ and $h_B$ are constant values within one packet); $n_R(t)$ is the zero-mean complex additive white Gaussian noise (AWGN) with double sided power spectral density $N_0/2$; $d_A$ and $d_B$ are the delays of signals $x_A(t)$ and $x_B(t)$ from nodes A and B to the relay R, and the symbol misalignment is given by $\tau=d_A-d_B$.

The received signal $r_R(t)$ is first passed through an RRC matched filter to get rid of the out-of-band noise, giving
\begin{eqnarray}\label{EquA4}
y_R(t) \!\!&=&\!\!  r_R(t)\ast p'(t)\nonumber\\
&=&\!\! h_A\sum_i s_A[i]p(t-iT-d_A)+h_B\sum_i s_B[i]p(t-iT-d_B)+ w_R(t),
\end{eqnarray}
where $\ast$ denotes the continuous-time convolution operation, $p(t)$ is the raised-cosine (RC) pulse (which is the effective pulse shape when one convolves two RRC pulses), $w_R(t)$ is the low-pass-filtered Gaussian noise term.

Given the continuous-time outputs of the RRC matched filter, we investigate two sampling schemes in this paper to obtain the discrete-time signal model.
\begin{itemize}
\item[(i)]
Baud-rate sampling. We sample $y_R(t)$ at baud-rate on equal-space positions $t=nT,~n=1,2,...$. The samples are given by
\begin{eqnarray}\label{EquA5}
y^b[n] \!\!&=&\!\!  y_R(t=nT)\nonumber\\
&=&\!\! h_A\sum_i s_A[i]p(nT-iT-t_A) + h_B\sum_i s_B[i]p(nT-iT-t_B)+ w^b[n]\nonumber\\
&\overset{(a)}{=}&\!\!h_A\sum_l s_A[n-l]p(lT-t_A) + h_B\sum_l s_B[n-l]p(lT-t_B)+ w^b[n],
\end{eqnarray}
where $t_A$ and $t_B$ are the time offsets with respect to the sampling positions\footnote{In~\eqref{EquA5}, we realign the time so that the sampling positions are with respect to the time kept by the receiver clock. Specifically, we sample $y_R(t)$ at $t=nT,n=1,2,...$, and the boundaries of symbols $s_A[n]$ and $s_B[n]$ with respect to the $n$-th sampling positions are now $t_A$ and $t_B$, respectively.} and the relative time offset $\tau=d_A-d_B=t_A-t_B$; step (a) follows by defining $l=n-i$; $\bm{w^b}$ is independent identically distributed (i.i.d.) Gaussian noise, $w^b[n]\sim\emph{CN}(0,\sigma^2)$ and $\sigma^2=N_0/2T$.
\item[(ii)]
Double baud-rate sampling. We sample $y_R(t)$ at double baud-rate on equal-space positions $t=kT/2,k=1,2,...$. The samples are given by
\begin{eqnarray}\label{EquA6}
y^d[k] \!\!&=&\!\!  y_R(t=\frac{k}{2}T)\nonumber\\
&=&\!\! h_A\sum_i s_A[i]p(\frac{k}{2}T-iT-t_A) + h_B\sum_i s_B[i]p(\frac{k}{2}T-iT-t_B)+ w^d[k],
\end{eqnarray}
\end{itemize}
where $t_A$ and $t_B$ are as defined in the baud-rate sampling case; $\bm{w^d}$ is colored noise, and each term $w^d[k]\sim\emph{CN}(0,\sigma^2)$.

We emphasize that the sampling process is a critical step in the RRC-APNC system, since it determines the amount of information we could obtain from the original continuous-time signal. When sampling the signals shaped by RC or RRC pulse, e.g., the RC pulse-shaped signal $y_R (t)$, baud-rate sampling is of low complexity but loses information when $\beta\neq 0$; while double baud-rate sampling guarantees no information-loss for all $\beta$ within the range $0\leq\beta\leq 1$.

The objective of relay R is to optimally decode the network-coded (XOR) information. To this end, two issues need to be addressed: (i) estimate the symbol misalignment accurately; (ii) devise an optimal decoder to find the most likely $s_{A\oplus B}[n]$ given by
\begin{eqnarray}\label{EquA7}
s_{A\oplus B}[n]=\arg\max_{s_{A\oplus B}[n]} \sum_{\substack{s_A[n],s_B[n]:\\s_A[n]\oplus s_B[n]=s_{A\oplus B}[n]}} \Pr\Big\{ s_A[n],s_B[n] \mid \bm{y^b}~\textup{or}~\bm{y^d}  \Big\}.
\end{eqnarray}

Estimating symbol misalignment is essentially estimate the unknown time offsets $t_A$ and $t_B$ in $\bm{y^b}$ and $\bm{y^d}$. This problem can be addressed by the application of two carefully designed preambles. These two preambles are assigned to nodes A and B, respectively, and are transmitted along with the data in each packet. At the receiver, we sample the overlapped preambles embedded in $y_R(t)$ and make use of the samples to estimate $t_A$ and $t_B$. The optimal estimators based on the two sampling schemes are detailed in Section III.

Given the estimated symbol misalignment and the samples corresponding to the data part, we investigate the design of optimal RRC-APNC decoders in the presence of inaccurate symbol misalignment estimates in Section IV. Specifically, we construct a factor graph with a tree structure, from which a sum-product algorithm (SPA) can be devised to optimally decode the network-coded messages following the metric in \eqref{EquA7}.
\section{Optimal Symbol-misalignment Estimators}
Estimating symbol misalignment is essentially a problem of estimating unknown parameters in the presence of random noise. An optimal estimator comes is the ML estimator~\cite{Proakisbook}. This section presents carefully designed preambles for symbol-misalignment estimation and derives the ML-optimal estimators under baud-rate sampling and double baud-rate sampling of the preambles.
\subsection{Preamble structure}
\begin{figure}[ht]
  \centering
  \includegraphics[width=0.5\columnwidth]{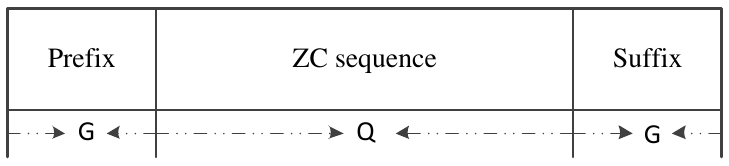}\\
  \caption{The structure of the preambles assigned to nodes A and B. It consists of a ZC sequence, a cyclic prefix and a cyclic suffix.}
\label{Fig2}
\end{figure}

The preambles assigned to nodes A and B have the same structure as shown in Fig.\ref{Fig2}. The preamble consists of a $Q$-points Zadoff-Chu (ZC) sequence along with a $G$-points cyclic prefix and a $G$-points cyclic suffix. ZC sequences are constant amplitude zero auto-correlation (CAZAC) sequences. They have a favorable auto-correlation property, that is, the auto-correlation of a ZC sequence with a cyclic-shifted version of itself is zero (i.e., any non-zero integer cyclic shift of the ZC sequence is orthogonal to the original ZC sequence). We assign cyclic-shifted ZC sequences to nodes A and B. Specifically, the ZC sequence assigned to node A (denote by $\bm{z_A}$) is a $\lfloor Q/2\rfloor$-points cyclic shifted version of that assigned to node B (denoted by $\bm{z_B}$). The cyclic prefix and cyclic suffix are to enable the computation of the cyclic cross-correlation in the presence of symbol misalignment.

This paper only considers fractional symbol misalignment, namely, $\tau\in[-T/2,T/2)$. As indicated in~\cite{QingEstTau}, the integral symbol misalignment can be estimated separately and it affects the performance of an ML estimator only when $E_b/N_0<0$ dB (i.e., there is essentially no error in the integral symbol misalignment estimate when $E_b/N_0\geq 0$ dB).
At the relay, the overlapping preambles are embedded in $y_R(t)$. Thus, we sample $y_R(t)$ at baud-rate or double baud-rate and utilize the samples corresponding to the preamble part to estimate the symbol misalignment. In the following, we first review the baud-rate estimator \cite{QingEstTau}, to get a preliminary understanding towards the ML estimation. Then, we present our double baud-rate estimator.
\subsection{Baud-rate estimator}
\begin{figure}[ht]
  \centering
  \includegraphics[width=0.7\columnwidth]{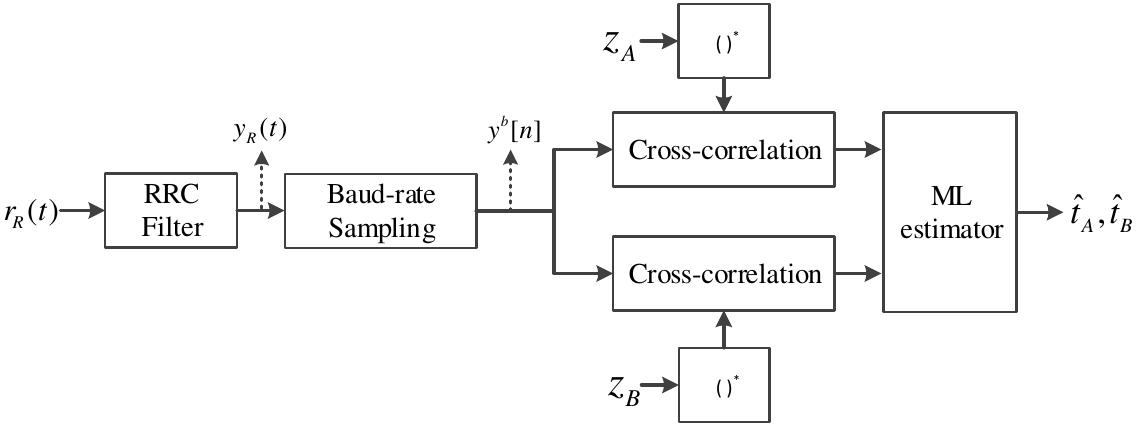}\\
  \caption{Signal processing flow for the symbol-misalignment estimation under baud-rate sampling.}
\label{Fig3}
\end{figure}

The signal processing flow of the baud-rate estimator is shown in Fig.\ref{Fig3}. As can be seen, given the baud-rate samples of $y_R(t)$ in \eqref{EquA5}, we cross-correlate $\bm{y^b}$ with the conjugate of the local $Q$-points ZC sequences, i.e. $\bm{z_A^\ast}$ and $\bm{z_B^\ast}$, respectively. Then, based on the cross-correlation outputs, the ML estimator can be designed.

In the following, we focus on the cross-correlation between $\bm{y^b}$ and $\bm{z_A^\ast}$, and the cross-correlation between $\bm{y^b}$ and $\bm{z_B^\ast}$ can be derived in the same way. The cross-correlation outputs are given by
\begin{eqnarray}\label{EquB1}
c^b_A[m]\!\!\!\!\!\!\!\!&&=\sum_{n=0}^{Q-1} z^{\ast}_A[n]y^b[n+m]\\
&&\!\!\!\!\!\!\!\!\!\!\!\!\!\!=h_A\sum_l p(lT\!-\!t_A)\sum_{n=0}^{Q-1}z_A^{\ast}[n]s_A[n\!+\!m\!-\!l]+h_B\sum_l p(lT\!-\!t_B)\sum_{n=0}^{Q-1}z_A^{\ast}[n]s_B[n\!+\!m\!-\!l]+\widetilde{w}^b_A[m],\nonumber
\end{eqnarray}
where $\widetilde{w}^b_A[m]=\sum_{n=0}^{Q-1}z_A^{\ast}[n]w^b[n+m]$; the transmitted preambles are contained in $\bm{s_A}$ and $\bm{s_B}$, respectively.

Thanks to the auto-correlation property of the ZC sequence, when the local $\bm{z_A^{\ast}}$ is aligned to the transmitted $\bm{z_A}$ embedded in $\bm{s_A}$, a peak emerges. Thus, we could detect the peak and locate the position of $\bm{z_A}$ in $\bm{s_A}$. In particular, the peak index is given by $I_A=\arg\max_m c^b_A[m]$. To ease exposition, we redefine the peak index as $0$ by offsetting $-I_A$ points if necessary, namely, we assume the peak emerges on $c_A^b[0]$.

In this context, if we only consider a small range $-d<m\leq d$ in $\bm{c_A^b}$, where $Q\gg d$, the second term on the R.H.S. of \eqref{EquB1} can be approximated as $0$. The reasons are as follows: when $m-l\neq\lfloor Q/2\rfloor$, the orthogonality of $\bm{z_A}$ and $\bm{z_B}$ makes it 0; when $m-l=\lfloor Q/2\rfloor$ (namely, $z_A[n]=z_B[n+m-l]$), $l$ must be large since $-d<m\leq d$, $Q\gg d$, and hence $p(lT-t_B)$ would be small so that this term can still be ignored. Thus, \eqref{EquB1} can be simplified as
\begin{eqnarray}\label{EquB2}
c^b_A[m]\!\!&\approx&\!\! h_A\sum_l p(lT-t_A)\sum_{n=0}^{Q-1}z_A^{\ast}[n]s_A[n+m-l]+\widetilde{w}^b_A[m]\nonumber\\
&\overset{(a)}{=}&\!\! h_A Q p(mT-t_A)+\widetilde{w}^b_A[m],
\end{eqnarray}
where $-d<m\leq d$; step (a) follows by retaining the term where $l=m$ only thanks to the orthogonality of $\bm{z_A}$ and its cyclic shift. Moreover, we specify that the auto-covariance function of the noise sequence $\bm{\widetilde{w}^b_A}$ is given by
\begin{eqnarray}\label{EquB3}
\mathbb{E}\Big[{\widetilde{w}^b_A[m]}^{\ast}\widetilde{w}^b_A[m+j]\Big]
\!\!&=&\!\!\mathbb{E}\Bigg[\sum_{n=0}^{Q-1}z_A[n]{w^b[n+m]}^{\ast}\sum_{n'=0}^{Q-1}z_A^{\ast}[n']w^b[n'+m+j]\Bigg]\nonumber\\
&=&\!\!\sigma^2\sum_{n=0}^{Q-1}\sum_{n'=0}^{Q-1}z_A[n]z_A^{\ast}[n']\delta(n'+j-n)\nonumber\\
&\overset{(a)}{=}&\!\!\sigma^2\sum_{n=\max\{0,j\}}^{\min\{Q-1,Q+j-1\}}z_A[n]z_A^{\ast}[n-j]\nonumber\\
&\overset{(b)}{\approx}&\!\!Q\sigma^2\delta(j),
\end{eqnarray}
where step (a) follows by setting $n'=n-j$, step (b) follows since $Q\gg j$.

The key implication from Eq.\eqref{EquB3} is that $\bm{\widetilde{w}^b_A}$ can be approximated as i.i.d. Gaussian noise, and $\widetilde{w}^b_A[m]\sim\emph{CN}(0,Q\sigma^2)$. Thus, we could directly perform ML estimation to estimate $t_A$ based on \eqref{EquB2}, because the ML metric will reduce to the minimum Euclidean distance metric \cite{Tsebook}.
The estimated $t_A$ is given by
\begin{eqnarray}\label{EquB4}
\widehat{t}_A\!\!&=&\!\!\arg\min_{t_A}\sum_{m=1-d}^{d}\left | c^b_A[m]-h_AQp(mT-t_A) \right |^2.
\end{eqnarray}

Similarly, we cross-correlate $\bm{y^b}$ and $\bm{z_B^\ast}$. Under the same assumption that $-d<m\leq d$, we have
\begin{eqnarray}\label{EquB5}
c^b_B[m]\!\!&\approx&\!\!h_B Q p(mT-t_B)+\widetilde{w}^b_B[m],
\end{eqnarray}
where $\bm{\widetilde{w}^b_B}$ can also be approximated as i.i.d. Gaussian noise and $\widetilde{w}^b_B[m]\sim\emph{CN}(0,Q\sigma^2)$. Thus, we could perform ML estimation to estimate $t_B$  based on \eqref{EquB5}, giving
\begin{eqnarray}\label{EquB6}
\widehat{t}_B\!\!&=&\!\!\arg\min_{t_B}\sum_{m=1-d}^{d}\left | c^b_B[m]-h_BQp(mT-t_B) \right |^2.
\end{eqnarray}

Finally, the estimated symbol misalignment is given by $\widehat{\tau}=\widehat{t}_A-\widehat{t}_B$.
\begin{rem}
The advantage of this baud-rate estimator is the low complexity compared with the oversampling scheme proposed in \cite{QingThesis} and \cite{DangEstTau}. However, for the signal shaped by the RC or RRC pulse, baud-rate is an undersampling rate when $\beta>0$. In other words, baud-rate is not sufficient to capture all the information contained in the original continuous-time signal relevant to estimating $t_A$ and $t_B$. A natural question is then whether the estimation can be further improved with a higher sampling rate? This motivates the design of our double baud-rate estimator in Subsection C.
\end{rem}

\subsection{Double baud-rate estimator}
According to the Nyquist-Shannon sampling theorem, the Nyquist rate is the lower bound on the sampling rate that can guarantee no information loss when converting the continuous-time signal into the discrete-time signal (and vice versa)~\cite{Proakisbook}. For the signals shaped by the RC or RRC pulse, the Nyquist rate is between baud-rate and double baud-rate (i.e., it is defined by a roll-off factor $\beta$ within the range of $0\leq\beta\leq 1$). This implies that double baud-rate sampling can capture all signal information embedded in $y_R(t)$  while averting unnecessary redundancy introduced by excess oversampling beyond double baud-rate sampling (e.g., 8-times sampling in \cite{DangEstTau}). Thus, in this subsection, we take the double baud-rate samples in \eqref{EquA6} as the inputs and devise an ML-optimal estimator.

\begin{figure}[ht]
  \centering
  \subfigure[Original processing flow (analog matched filtering)]
  {
  \label{Fig4a}
  \includegraphics[width=0.8\columnwidth]{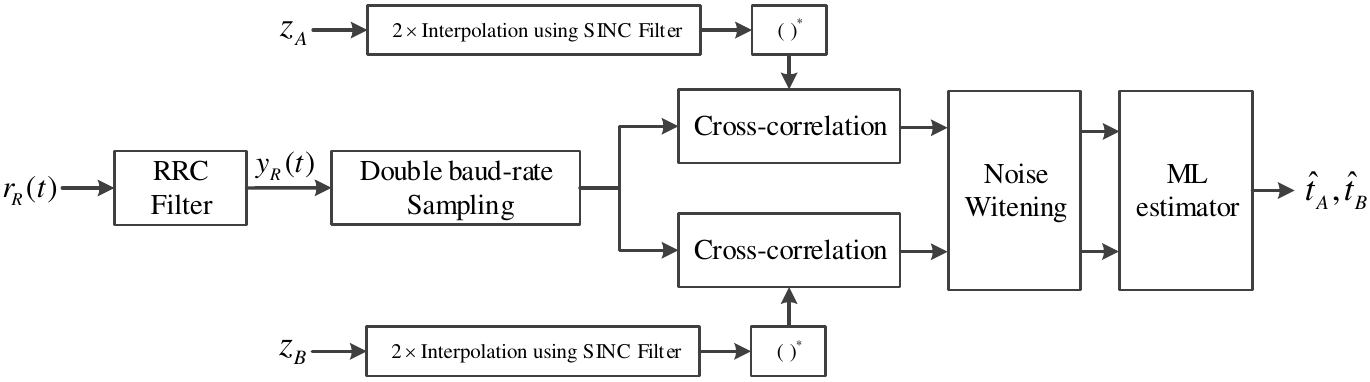}
  }
  \subfigure[Equivalent processing flow (digital matched filtering)]
  {
  \label{Fig4b}
  \includegraphics[width=0.8\columnwidth]{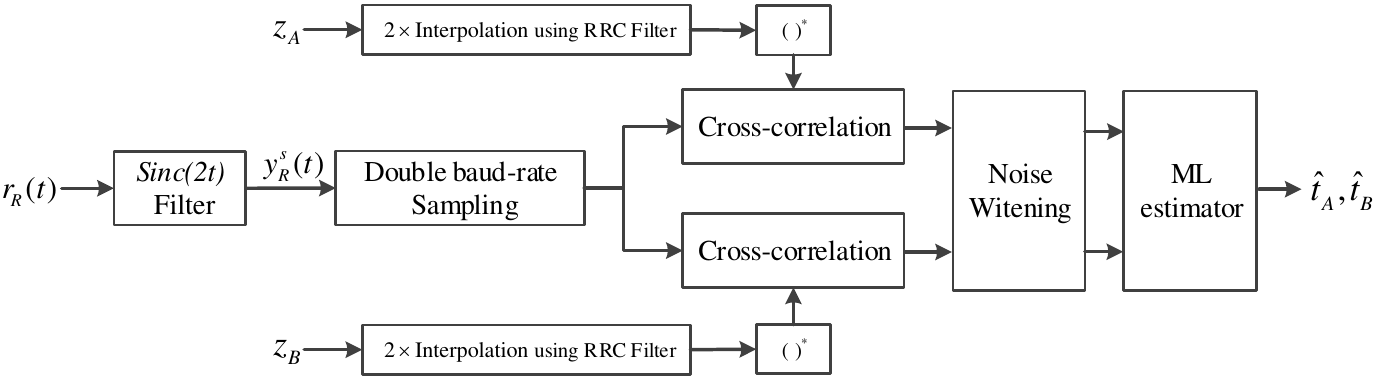}
  }
  \caption{Signal processing flows for the symbol-misalignment estimation under double-baud-rate sampling.}
  \label{Fig4}
\end{figure}

The signal processing flow of our double baud-rate estimator is shown in Fig.\ref{Fig4a}, where the double baud-rate samples will first be cross-correlated with the local ``interpolated double baud-rate ZC sequences''. A continuous-time ZC sequence is the result of modulating an original discrete-time ZC sequence using a certain continuous-time pulse, e.g., the $Sinc(t)$ pulse in Fig.\ref{Fig4a}. The ``2$\times$ interpolation'' means the resulting discrete-time ZC sequence after sampling the continuous-time ZC sequence at double baud-rate. Thus, the ``interpolated double baud-rate ZC sequence'' has twice the samples of the original ZC sequence (basically, it is the original sequence with a zero inserted between two ZC samples given than $Sinc(t)$ is assumed).

A new difficulty in the analysis of Fig.\ref{Fig4a} is that, the noise term in the double baud-rate samples, i.e., $\bm{w^d}$ in \eqref{EquA6}, is no longer white. To facilitate analysis, we put forth an equivalent signal processing scheme as shown in Fig.\ref{Fig4b}. Two distinctions between the flow diagrams in Fig.\ref{Fig4a} and Fig.\ref{Fig4b} are as follows:
\begin{itemize}
\item[(i)] The RRC matched filter on the front-end of Fig.\ref{Fig4a} is replaced with a $Sinc(2t)$ filter in Fig.\ref{Fig4b}. In this case, the noise terms in the samples would be white, and the analyses afterwards could be simplified.
\item[(ii)] In Fig.\ref{Fig4b}, given the $Sinc(2t)$ filter on the front-end, the noise power in each sample is doubled (i.e., $2\sigma^2$). Thus, a digital RRC filter is needed to filter the out-of-band noise. In Fig.\ref{Fig4b}, the ZC sequences $\bm{z_A}$  and $\bm{z_B}$  are twice interpolated using the RRC filter rather than the $Sinc(t)$ filter in Fig.\ref{Fig4a}.\footnote{Accordingly, an interpolated sample between two original ZC samples are nonzero in Fig.\ref{Fig4b} due to the use of RRC interpolation rather than the SINC interpolation in Fig.\ref{Fig4a}.} This implies that the cross-correlation operation in Fig.\ref{Fig4b} also embeds a process of digital matched filtering (i.e., the analog matched filtering in Fig.\ref{Fig4a} is moved to the digital domain in Fig.\ref{Fig4b}).
\end{itemize}
Essentially, the transformation from Fig.\ref{Fig4a} to Fig.\ref{Fig4b} is a conversion from analog matched filtering to digital matched filtering. Thus, these two flows are equivalent. In particular, after cross-correlation, the signals from the two flows are identical.

In the following, we detail the signal processing flow in Fig.\ref{Fig4b} step by step. Specifically, we explain how to utilize the double baud-rate samples to obtain more accurate estimation of $\tau$.

\subsubsection{Sampling}
The received signal $r_R(t)$ is first fed to a $Sinc(2t)$ filter, giving
\begin{eqnarray}\label{EquB7}
y^s_R(t) \!\!&=&\!\!  r_R(t)\ast {Sinc}(2t)\nonumber\\
&=&\!\! h_A\sum_i s_A[i]p'(t-iT-d_A)+h_B\sum_i s_B[i]p'(t-iT-d_B)+ w^s_R(t),
\end{eqnarray}
where $w^s_R(t)$ is the low-pass-filtered Gaussian noise term and the power of $w^s_R(t)$ is doubled compared with that of $w_R (t)$ in \eqref{EquA4}.

Next, we sample \eqref{EquB7} at double baud-rate, and the samples are given by
\begin{eqnarray}\label{EquB8}
\bar{y}^d[k] \!\!&=&\!\!  y^s_R(t=\frac{k}{2}T)\nonumber\\
&=&\!\! h_A\sum_i s_A[i]p'(\frac{k}{2}T-iT-t_A)+h_B\sum_i s_B[i]p'(\frac{k}{2}T-iT-t_B)+ \bar{w}^d[k],
\end{eqnarray}
where $\bm{\bar{w}^d}$ is now i.i.d. Gaussian noise, $\bar{w}^d[k]\sim CN(0,2\sigma^2 )$.

Meanwhile, we ``double-interpolate'' the local ZC sequences $\bm{z_A, z_B}$ using the RRC filter. The interpolated ZC sequences, denoted by $\bm{z^d_A, z^d_B}$, are given by
\begin{eqnarray}
\label{EquB9}
z^d_A[k] \!\!&=& \!\! \sum_{i=0}^{Q-1} z_A[i]p'(\frac{k}{2}T-iT),\\
\label{EquB10}
z^d_B[k] \!\!&=& \!\! \sum_{i=0}^{Q-1} z_B[i]p'(\frac{k}{2}T-iT),
\end{eqnarray}

\subsubsection{Cross-correlation}
Next, we cross-correlate $\bm{{z^d_A}^*,{z^d_B}^*}$ with $\bm{\bar{y}^d}$, respectively.
Focusing on the the cross-correlation between $\bm{{z^d_A}^*}$ and $\bm{\bar{y}^d}$, we have
\begin{eqnarray}\label{EquB11}
c^d_A[m]\!\!\!\!\!\!\!\!&&=\sum_{k=0}^{2Q-1}{z^d_A}^*[k]\bar{y}^d[k+m]\nonumber\\
&&\overset{(a)}{=}Qh_A\sum_l\Big[p'(lT)p'(\frac{m}{2}T+lT-t_A)+p'(lT+\frac{T}{2})p'(\frac{m+1}{2}T+lT-t_A)\Big] +\widetilde{w}^d_A[m]\nonumber\\
&&\overset{\textup{def}}{=} v^d_A[m]+\widetilde{w}^d_A[m],
\end{eqnarray}
where step (a) involves a number of steps, the details of which are given in Appendix A.
We further prove in Appendix A that, the noise term $\bm{\widetilde{w}^d_A}$ is colored overall, but its odd and even entries are white among themselves.
As stated above, the outcomes of cross-correlation in Fig.\ref{Fig4a} and Fig.\ref{Fig4b} are identical. Thus, \eqref{EquB11} is also the cross-correlation outputs in Fig.\ref{Fig4a}.

If we directly perform ML estimation on \eqref{EquB11}, the ML metric would be quiet complicated and there may be no analytical solution owing to the colored noise. Thus, we will first whiten the noise, and then construct the ML metric based on the whitened sequence.

\subsubsection{Noise whitening}
To whiten the noise term $\bm{\widetilde{w}^d_A}$, it is important to study its covariance matrix $\bm{\Sigma(\widetilde{w}^d_A)}$, since noise whitening is a process that transform $\bm{\Sigma(\widetilde{w}^d_A)}$ to a diagonal matrix. We show in Appendix A that, if we only consider a small range $-d<m\leq d$ ($Q\gg d$), the auto-covariance function of $\bm{\widetilde{w}^d_A}$ can be expressed as
\begin{eqnarray}\label{EquB12}
\mathbb{E}\big[{\widetilde{w}^d_A[m]}^*\widetilde{w}^d_A[m+j] \big]\approx 2Q\sigma^2 p(\frac{j}{2}T),
\end{eqnarray}
where the approximation comes from the infinite time continuation of the RRC pulse and the limited prefix/suffix length.

Given \eqref{EquB12}, the covariance matrix of $\bm{\widetilde{w}^d_A}$ can be written as
\begin{eqnarray}\label{EquB13}
\bm{\Sigma(\widetilde{w}^d_A)} = 2Q\sigma^2\bm{\Sigma_0}
=2Q\sigma^2
\begin{bmatrix}
\begin{smallmatrix}
1                   & p(\frac{1}{2}T)   & 0                     & p(\frac{3}{2}T)   & \cdots & p(\frac{2d-1}{2}T)   \\[0.2cm]
p(-\frac{1}{2}T)    & 1                 & p(\frac{1}{2}T)       & 0                 & \cdots & 0                    \\[0.2cm]
0                   & p(-\frac{1}{2}T)  & 1                     & p(\frac{1}{2}T)   & \cdots & p(\frac{2d-3}{2}T)    \\[0.2cm]
p(-\frac{3}{2}T)    & 0                 & p(-\frac{1}{2}T)      & 1                 & \cdots & 0                    \\[0.2cm]
\vdots              & \vdots            & \vdots                & \vdots            & \ddots & \vdots               \\[0.2cm]
p(\frac{1-2d}{2}T)  & 0                 & p(\frac{3-2d}{2}T)    & 0                 & \cdots & 1
\end{smallmatrix}
\end{bmatrix}.
\end{eqnarray}
Thus, if we could transform the matrix $\bm{\Sigma_0}$ to a diagonal matrix, the noise is whitened naturally.

A possible transformation is to use Cholesky decomposition. Cholesky decomposition makes use of the fact that any real-valued symmetric positive-definite matrix can be factorized into the product of a lower triangular matrix and its transpose \cite{Matrix}. In the following, we show first why $\bm{\Sigma_0}$ can be Cholesky decomposed, then how the Cholesky decomposition can help whiten the noise.

First, it is obvious that $\bm{\Sigma_0}$ is real and symmetric since $p(t)$ is an even function. Meanwhile, we prove in Appendix B that $\bm{\Sigma_0}$ is positive definite. Thus, $\bm{\Sigma_0}$ can be Cholesky decomposed and we have $\bm{\Sigma_0}=\bm{U_0}^T\bm{U_0}$, where $\bm{U_0}$ is an invertible upper triangular matrix (its diagonal entries are nonzero).

Then, considering the small range $-d<m\leq d$ ($Q\gg d$), we could rewrite \eqref{EquB11} in vector form as
\begin{eqnarray}\label{EquB14}
\bm{c^d_A=v^d_A+\widetilde{w}^d_A}
\end{eqnarray}
where $\bm{c^d_A}=\big[c^d_A[1\!-\!d],...,c^d_A[d]\big]^T$, $\bm{v^d_A}=\big[v^d_A[1\!-\!d],...,v^d_A[d]\big]^T$ and $\bm{\widetilde{w}^d_A}=\big[\widetilde{w}^d_A[1\!-\!d],...,\widetilde{w}^d_A[d]\big]^T$.
Next, we left multiply \eqref{EquB14} by $\bm{U_0}^{-T}$, giving
\begin{eqnarray}\label{EquB15}
\bm{{U_0}^{-T}c^d_A={U_0}^{-T}v^d_A+{U_0}^{-T}\widetilde{w}^d_A}~~~~~~\overset{\textup{def}}{=}~~~~~~\ddot{\bm{c_A}}=\ddot{\bm{v_A}}+\ddot{\bm{w_A}}.
\end{eqnarray}
The transformed noise $\ddot{\bm{w_A}}$ in \eqref{EquB15} is now i.i.d. Gaussian noise since $\mathbb{E}\big[\ddot{\bm{w_A}}\ddot{\bm{w_A}}^H\big]$ $=$ $2Q\sigma^2$ $\bm{U_0}^{-T}\bm{\Sigma_0}\bm{U_0}^{-1}$ $=$ $2Q\sigma^2\bm{I}$, where $\bm{I}$ is the identity matrix.
Thus, \eqref{EquB15} is exactly the whitened discrete-time model corresponding to \eqref{EquB14}.

Similarly, we cross-correlate $\bm{{z^d_B}^*}$ with $\bm{\bar{y}^d}$ and whiten the cross-correlation results. The corresponding whitened model can be written as
\begin{eqnarray}\label{EquB16}
\ddot{\bm{c_B}}=\ddot{\bm{v_B}}+\ddot{\bm{w_B}},
\end{eqnarray}
where $\ddot{\bm{c_B}}$, $\ddot{\bm{v_B}}$ and $\ddot{\bm{w_B}}$ are the counterparts of $\ddot{\bm{c_A}}$, $\ddot{\bm{v_A}}$ and $\ddot{\bm{w_A}}$ in \eqref{EquB15}, respectively.

\subsubsection{ML estimation}
Finally, based on the whitened model in \eqref{EquB15} and \eqref{EquB16}, we could directly construct the metrics for ML estimation as follows:
\begin{eqnarray}
\label{EquB17}
\widehat{t}_A\!\!&=&\!\!\arg\min_{t_A}\mid\ddot{\bm{c_A}}-\ddot{\bm{v_A}} \mid^2,\\
\label{EquB18}
\widehat{t}_B\!\!&=&\!\!\arg\min_{t_B}\mid\ddot{\bm{c_B}}-\ddot{\bm{v_B}} \mid^2.
\end{eqnarray}

The estimated symbol misalignment is given by $\widehat{\tau}=\widehat{t}_A-\widehat{t}_B$.

\subsection{Numerical results}
This subsection evaluates the performance of our double baud-rate estimator benchmarked against a baud-rate estimator. Let $\widehat{\tau}^b$ and $\widehat{\tau}^d$ be the estimated symbol misalignments using baud-rate and double baud-rate estimators. Their respective square errors are
\begin{eqnarray}
\label{EquB19}
\varepsilon^b\!\!&=&\!\!(\widehat{\tau}^b-\tau)^2,\\
\label{EquB20}
\varepsilon^d\!\!&=&\!\!(\widehat{\tau}^d-\tau)^2,
\end{eqnarray}
where $\tau$ is the real symbol misalignment.
In particular, we are interested in the the mean-square-error (MSE) and the probability density function (PDF) of the square error.

\begin{figure}[ht]
  \centering
  \includegraphics[width=0.6\columnwidth]{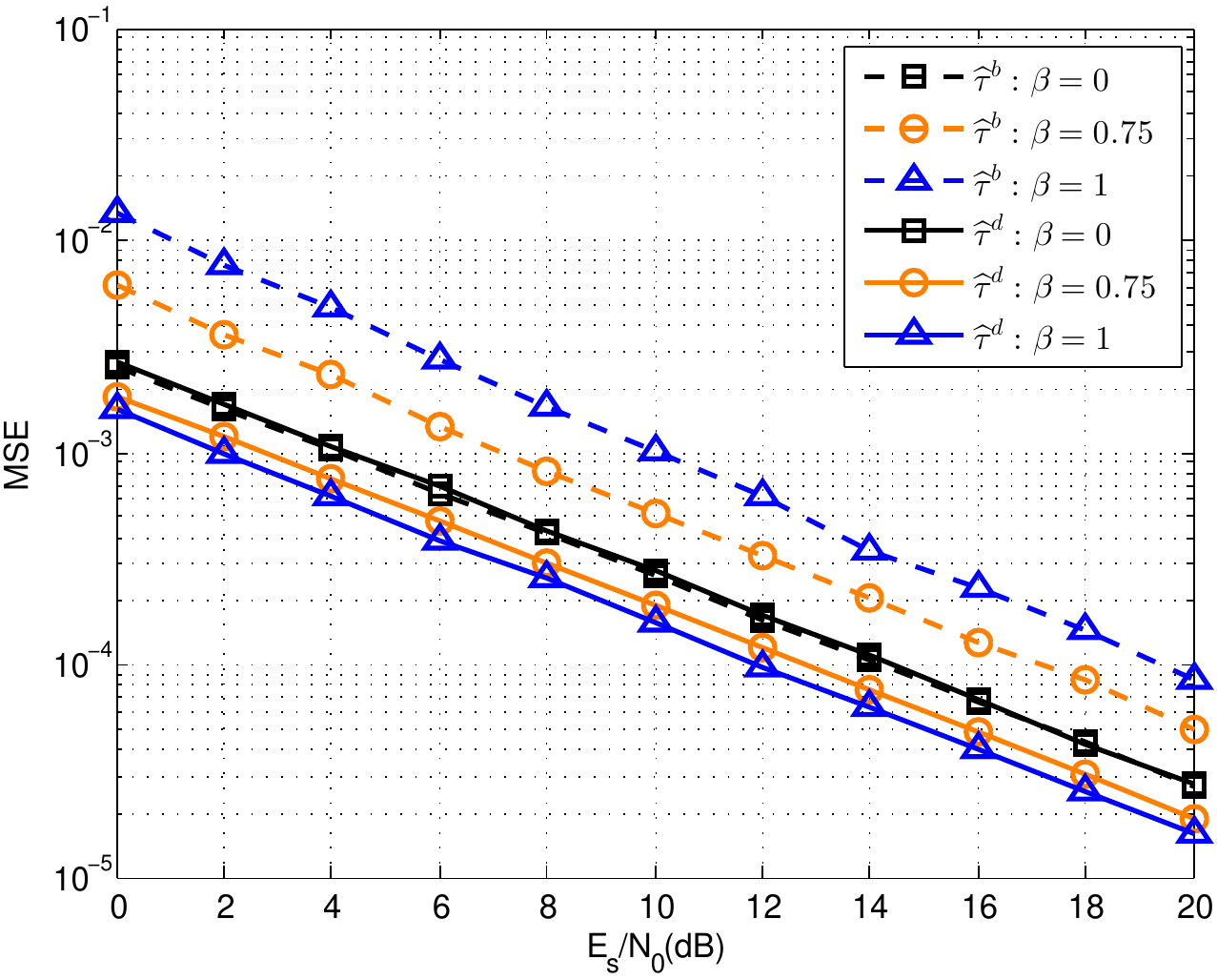}\\
  \caption{The MSEs of $\widehat{\tau}^b$  and $\widehat{\tau}^d$ under different $\beta$ and fixed $Q$. We set $Q=31,~\beta=0,~0.75~\textup{or}~1,~G=10,~d=4,~T=1,~h_A=h_B=1$, $\tau$ is uniformly distributed in $[-0.5,0.5)$. The dashed curves mark the MSE of $\widehat{\tau}^b$ while the solid curves mark the MSE of $\widehat{\tau}^d$.}
\label{FigSim1}
\end{figure}

\subsubsection{MSE Performance}
First, we set $h_A=h_B=1$ (i.e., AWGN channel), and evaluate the MSE of $\widehat{\tau}^d$ given a fixed ZC-sequence length $Q=31$ and different roll-off factors $\beta=0,~0.75~\textup{or}~1$.

The simulation results are shown in Fig.\ref{FigSim1}. When $\beta=0$ (the case of SINC pulse), the two curves that mark the MSEs of $\widehat{\tau}^b$ and $\widehat{\tau}^d$ coincide with each other.\footnote{In fact, from Fig.\ref{FigSim1}, the MSE of $\widehat{\tau}^b$  is slightly better than that of $\widehat{\tau}^d$  when $\beta=0$. The reason is that the whitening process is sensitive to the assumption $Q\gg j$. If we further increase the length of ZC sequence, e.g. we set Q=63, d=4, these two curves would coincide exactly.} This outcome is expected since baud-rate is already the Nyquist rate for SINC pulse. However, with the increase of $\beta$, the performance of $\widehat{\tau}^b$ degrades since the baud-rate is an undersampling rate when $0<\beta\leq 1$; whereas for $\widehat{\tau}^d$, the performance improves with the increase of $\beta$.
Overall, the performance of $\widehat{\tau}^d$  is superior to that of $\widehat{\tau}^b$. The performance gain of $\widehat{\tau}^d$  over $\widehat{\tau}^b$ increases with the increase of $\beta$. When $\beta=1$, the performance gap is about $8$ dB.
\begin{figure}[ht]
  \centering
  \includegraphics[width=0.6\columnwidth]{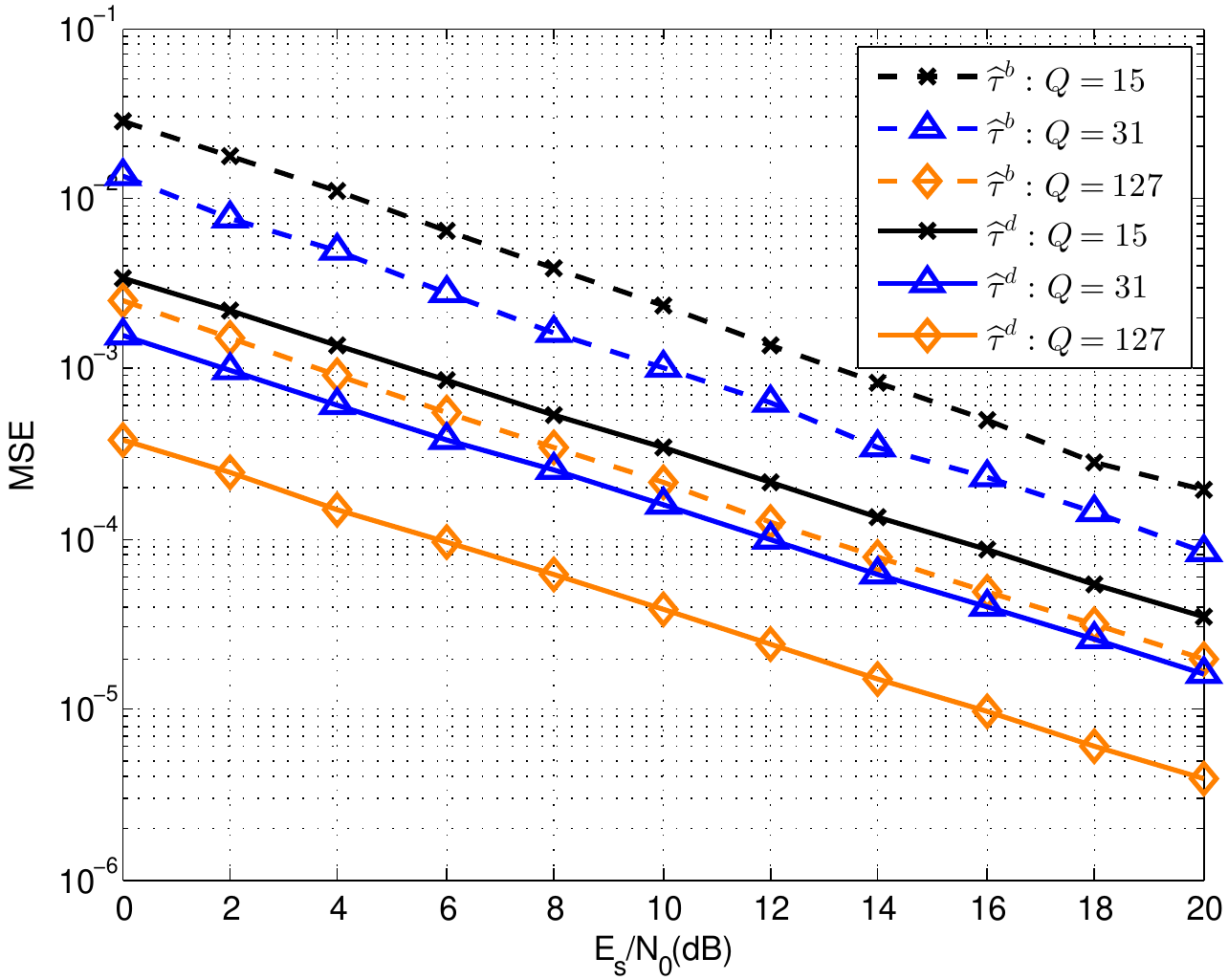}\\
  \caption{The MSEs of $\widehat{\tau}^b$  and $\widehat{\tau}^d$ under different $Q$ and fixed $\beta$. We set $\beta=1,~Q=15,~31~\textup{or}~127,~G=\lfloor Q/3\rfloor,~d=4,~T=1,~h_A=h_B=1$, $\tau$ is uniformly distributed in $[-0.5,0.5)$. The dashed curves mark the MSE of $\widehat{\tau}^b$ while the solid curves mark the MSE of $\widehat{\tau}^d$.}
\label{FigSim2}
\end{figure}

Then, given a fixed roll-off factor $\beta=1$ and different ZC-sequence length $Q=15,~31~\textup{or}~127$, we evaluate the MSEs of $\widehat{\tau}^b$ and $\widehat{\tau}^d$ in the AWGN channel.
The simulation results are shown in Fig.\ref{FigSim2}. As can be seen, the performance of both estimators improves with the increase of $Q$, and the double baud-rate estimator consistently outperforms the baud-rate estimator.
\begin{figure}[ht]
  \centering
  \includegraphics[width=0.6\columnwidth]{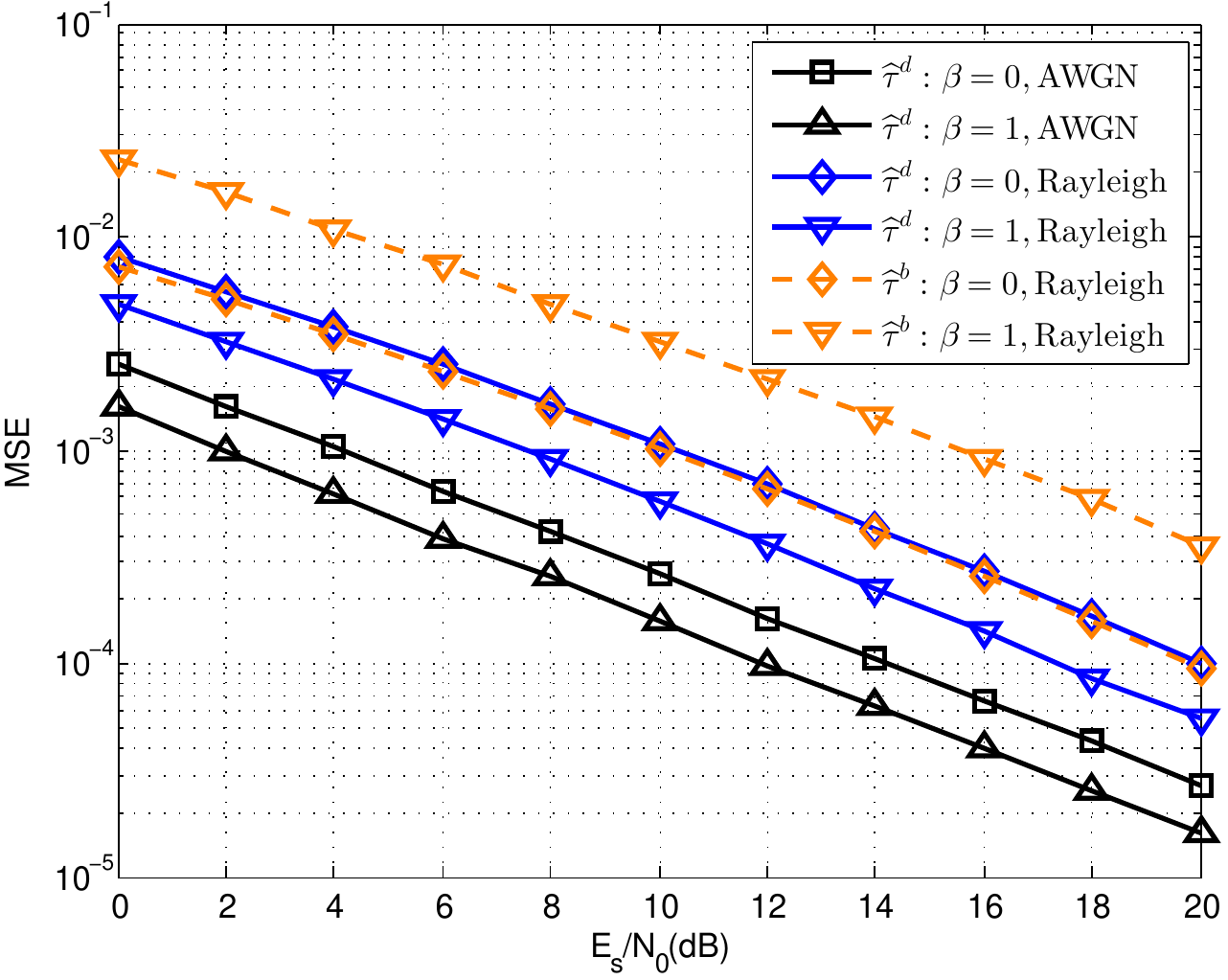}\\
  \caption{The MSEs of $\widehat{\tau}^b$  and $\widehat{\tau}^d$ in a flat Rayleigh fading channel. we set $Q=31,~\beta=0~\textup{or}~1,~G=10,~d=4,~T=1$, the real relative time offset $\tau$ is assumed to be uniformly distributed in $[-0.5,0.5)$; and the channel gains $h_A$ and $h_B$ are Rayleigh distributed.}
\label{FigSim3}
\end{figure}

Fig.\ref{FigSim3} presents the MSEs of $\widehat{\tau}^b$ and $\widehat{\tau}^d$ in a flat Rayleigh fading channel.
After averaging over a large number of $(h_A, h_B)$ pairs, it can be seen that,
(i) the conclusions drawn in the AWGN channel (see Fig.\ref{FigSim1}) still holds in the flat Rayleigh fading channel. Specifically, the gain of the double baud-rate estimator over the baud-rate estimator is also about $8$ dB when $\beta=1$ under Rayleigh fading.
(ii) compared with AWGN channel, the Rayleigh fading channel leads to a $6$ dB degradation of the MSEs for both $\widehat{\tau}^b$ and $\widehat{\tau}^d$.
However, it is worth noting that MSE does not fully characterize the estimation performance, since it is only a first-order statistical characteristic of square error. Thus, we further analyze the PDF of the square error in the following.

\subsubsection{PDF of Square Error}

\begin{figure}[ht]
  \centering
  \subfigure[AWGN channel]
  {
  \label{FigSim4a}
  \includegraphics[width=0.46\columnwidth]{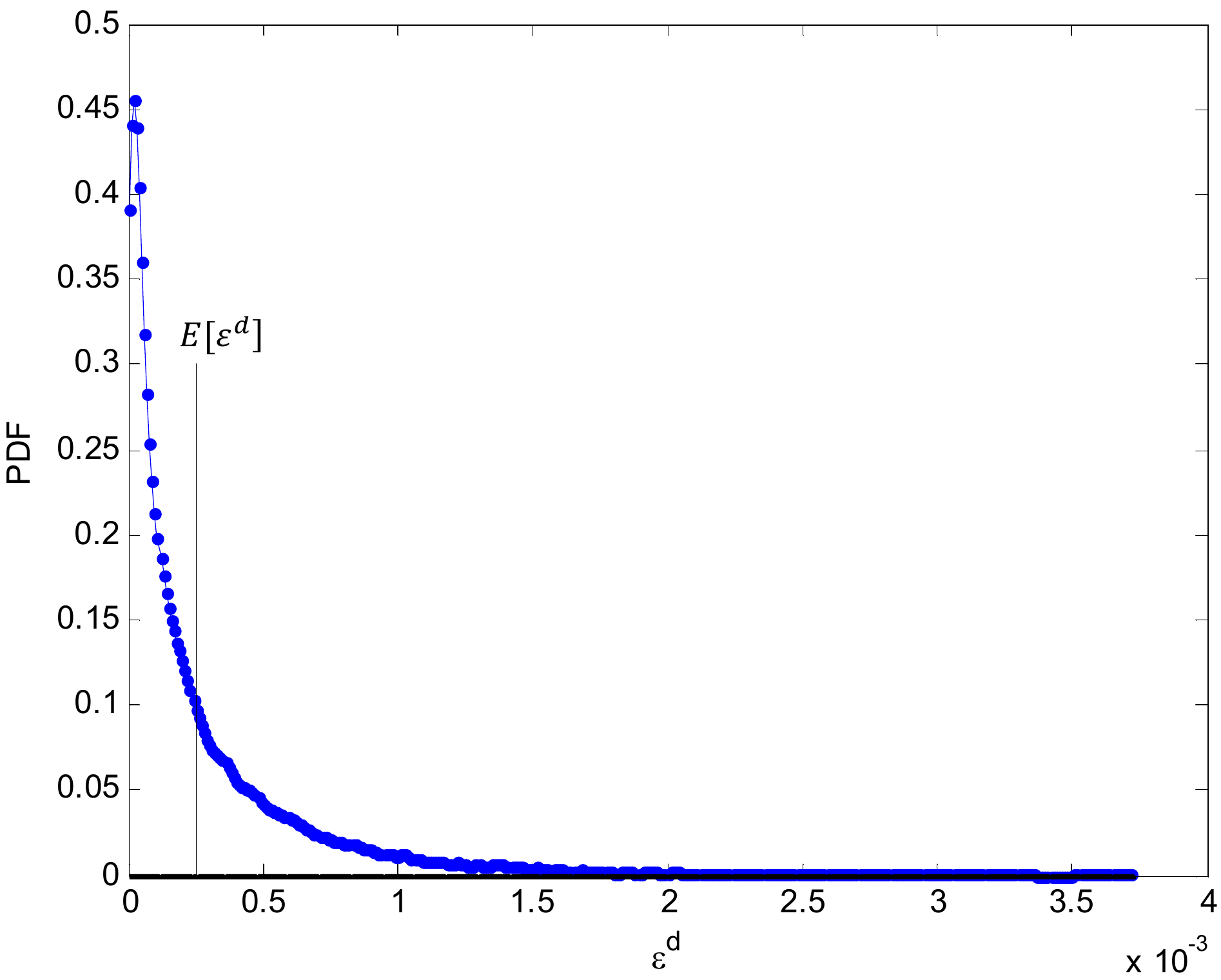}
  }
  \subfigure[Rayleigh fading channel]
  {
  \label{FigSim4b}
  \includegraphics[width=0.46\columnwidth]{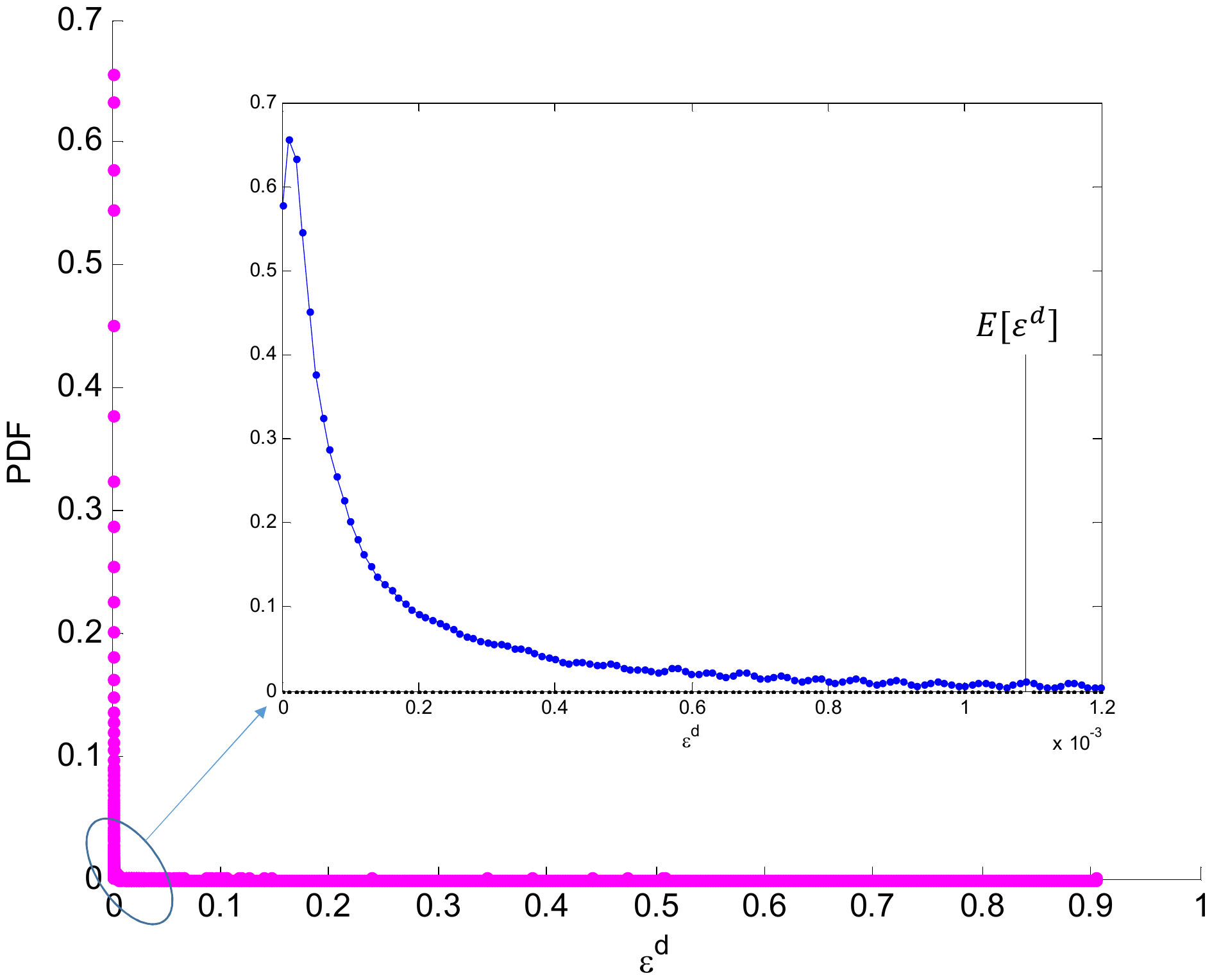}
  }
  \caption{The PDF of $\varepsilon^d$ in the AWGN and flat Rayleigh fading channels. We set $Q=31,~\beta=0,~E_b/N_0=10,~G=10,~d=4,~T=1$, $\tau$ is uniformly distributed in $[-0.5,0.5)$. In Fig.\ref{FigSim4a}, $h_A=h_B=1$. In Fig.\ref{FigSim4b}, $h_A$ and $h_B$ are Rayleigh distributed.}
  \label{FigSim4}
\end{figure}

The PDFs of $\varepsilon^d$ in the AWGN and flat Rayleigh fading channels are shown in Fig.\ref{FigSim4a} and Fig.\ref{FigSim4b}, respectively, where we fix $Q=31$, $\beta=0$,\footnote{Note that $\beta=0$ is the worst case for our double baud-rate estimator.} $E_b/N_0=10$ dB. Given these two figures, we have the following two key observations:
\begin{itemize}
\item In the AWGN channel, $\varepsilon^d\in[0,~0.004)$; while in the flat Rayleigh fading channel, $\varepsilon^d\in[0,1)$. This is intuitive since the estimation may be totally off when Rayleigh channel is under deep fading.
\item In both channels, $\varepsilon^d$ falls into the range $[0,~0.001]$ with high probability, even though the MSEs differ greatly.
\end{itemize}

Recall that in Fig.\ref{FigSim3}, the MSE performance of our double baud-rate estimator suffers a $6$ dB degradation in Rayleigh channel compared with that in an AWGN channel.
However, the two observations above imply that, the large square error under deep fading in the Rayleigh fading channel may pull down the average performance (i.e., the MSE performance). The real performance of our double baud-rate estimator in a Rayleigh fading channel is comparable to that in AWGN channel.
If we define good/acceptable estimation, say, to be $\varepsilon^d\leq 0.001$, the probability that a good estimation occurs in AWGN channel is about $94.69\%$, while that in Rayleigh channel is about $90.39\%$.

\section{Optimal RRC-APNC Decoders}
Given the estimated symbol misalignment in Section III, this section devises optimal RRC-APNC decoders assuming baud-rate and double baud-rate sampling on the data part. As stated in the system model, an optimal RRC-APNC decoder aims to find the most likely network-coded (XOR) symbol $s_{A\oplus B}[n]$ in accordance with the metric in \eqref{EquA7}.

For a conventional point-to-point system with RRC pulse, the optimal decoder is simply a symbol-by-symbol decoder since we could always sample at specific positions to eliminate the ISI \cite{Proakisbook}. However, for RRC-APNC, when the symbol boundaries of A and B are not aligned, ISI is inevitable because there are no ISI-free sampling positions for both signals from nodes A and B.

Under ISI, an ideal optimal PNC decoder should decode $s_{A\oplus B}[n]$ based on all the samples within the packet since the information about each $s_{A\oplus B}[n]$ is spread across the whole packet given the long tails of the RRC pulse. This gives rise to a critical issue: the enormous state size of the decoder. Specifically, if the data length in a single packet is $N$, then each sample in \eqref{EquA5} or \eqref{EquA6} is related to $2N$ transmitted symbols from both nodes A and B, and each observation node\footnote{The terms ``observation node'', ``variable node'' and ``check node'' in this paper follow the conventional definitions in \cite{SPA,BP}. In particular, an observation node corresponding to a single sampling point.} of an optimal decoder will have $q^{2N}$ states ($q$ is the modulation order). Given the state size grows exponentially with $N$, the ideal optimal PNC decoder is infeasible when complexity is taken into account.
In this context, this paper takes the following two measures to enable feasible decoders:

\begin{itemize}
\item[(i)] Sample at the symbol boundaries -- As indicated in \eqref{EquA5} or \eqref{EquA6}, the samples at the symbol boundaries of $\bm{s_A}$  or $\bm{s_B}$ (i.e. the samples at $nT+t_A$ or $nT+t_B$, $n=1,2,...,N$) contain no information relevant to $s_A[n'],~n'\neq n$ or $s_B [n],~n'\neq n$ since RC pulse meets the Nyquist ISI-free criterion. In other words, we could eliminate the ISI from one user if we sample at the symbol boundaries of this user.
\item[(ii)] Truncate states -- Measure $(i)$ can eliminate the ISI of one user only; the long-tailed ISI of the other user remains. Practically, given a fixed processing capacity of the receiver -- e.g., the size of states that the receiver can handle is at most $q^L$ -- the only choice is to make an approximation by truncating the states.
\end{itemize}

Based on the two measures above, this paper constructs a factor graph, for which a sum-product algorithm (SPA) \cite{SPA,BP} can be used to decode the network-coded symbols. This factor graph has a tree structure, and hence iterative decoding is not necessary inside the PNC decoder. For a channel-coded system, however, we will have an additional factor graph corresponding to the channel code. These two factor graphs are combined into one for an overall system and turbo decoding is also possible \cite{Turbo}.
In the following, we focus on the design of the double baud-rate decoder. The design of the baud-rate decoder is elaborated in Appendix C.

\subsection{Discrete-time model}
We have obtained the double baud-rate samples of $y_R(t)$ in \eqref{EquA6}. The estimated time offsets of signals from two end nodes have been given in Section III. Thus, we first follow the measure (i) above and derive the samples at the symbol boundaries of $\bm{s_A}$ and $\bm{s_B}$.

Double baud-rate sampling is information-lossless when converting $y_R(t)$ to the discrete-time signal. Thus, the double baud-rate samples $\bm{y^d}$ can be used to reconstruct the continuous-time signal $y_R(t)$. That is,
\begin{eqnarray}\label{EquC1}
\widehat{y}_R(t)=\sum_k y^d[k]Sinc(2t-kT),
\end{eqnarray}
As indicated by the Nyquist-Shannon sampling theorem \cite{Proakisbook}, we have $\widehat{y}_R(t)=y_R(t)$.

Next, given the estimated time offsets $\widehat{t}_A$ and $\widehat{t}_B$, the samples at the symbol boundaries (i.e., $nT+\widehat{t}_A$ and $nT+\widehat{t}_B$) can be obtained by (the following computation can be interpreted as a ``virtual resampling process'' whereby the samples at the two boundaries are obtained from $\widehat{y}_R(t)$ in \eqref{EquC1} through computation rather than actual resampling)
\begin{eqnarray}
\label{EquC2}
y_1[n]\!\!&=&\!\!\widehat{y}_R(t=nT+\widehat{t}_A)\nonumber\\
&=&\!\!h_A\sum_l s_A[n-l]p(lT+\widehat{t}_A-t_A)+h_B\sum_l s_B[n-l]p(lT+\widehat{t}_A-t_B)+w_1[n];\\
\label{EquC3}
y_2[n]\!\!&=&\!\!\widehat{y}_R(t=nT+\widehat{t}_B)\nonumber\\
&=&\!\!h_A\sum_l s_A[n-l]p(lT+\widehat{t}_B-t_A)+h_B\sum_l s_B[n-l]p(lT+\widehat{t}_B-t_B)+w_2[n],
\end{eqnarray}
where $\bm{w_1}$ and $\bm{w_2}$ are i.i.d. Gaussian noise, $w_1[n],w_2[n]\sim\emph{CN}(0,\sigma^2)$, but $[\bm{w_1,~w_2}]^T$ is colored because a component in $\bm{w_1}$ may not be independent of a component in $\bm{w_2}$. Notice that if the estimated boundaries are exact (i.e., $\widehat{t}_A=t_A$ and $\widehat{t}_B=t_B$), the samples $y_1[n]$ and $y_2[n]$ contain information only relevant to the variables $\{s_A[n],~s_B[n'],~n'=1,2,...,N\}$ and $\{s_B[n],~s_A[n'],~n'=1,2,...,N\}$, respectively.

Further, we can combine \eqref{EquC2} and \eqref{EquC3} and rewrite them in matrix form:
\begin{eqnarray}\label{EquC4}
\bm{y=\Gamma H x + w},
\end{eqnarray}
where the vectors $\bm{y}$, $\bm{x}$, $\bm{w}$ and the matrix $\bm{H}$ are defined as
\begin{eqnarray}
\label{EquC5}
\bm{y}\!\!&=&\!\!\Big[y_1[1],y_2[1],~y_1[2],y_2[2],~...,...,~y_1[N],y_2[N],\Big]^T,\\
\label{EquC6}
\bm{x}\!\!&=&\!\!\Big[s_A[1],s_B[1],~s_A[2],s_B[2],~...,...,~s_A[N],s_B[N],\Big]^T,\\
\label{EquC7}
\bm{w}\!\!&=&\!\!\Big[w_1[1],w_2[1],~w_1[2],w_2[2],~...,...,~w_1[N],w_2[N],\Big]^T,\\
\label{EquC8}
\bm{H}\!\!&=&\!\!\textup{diag}(h_A,h_B,h_A,h_B,...,...,h_A,h_B).
\end{eqnarray}
The coefficient matrix $\bm{\Gamma}$ is given by
\begin{eqnarray}\label{EquC9}
\bm{\Gamma} =
\begin{bmatrix}
\begin{smallmatrix}
p(\widehat{t}_A-t_A)& p(\widehat{t}_A-t_B)&p(\widehat{t}_A-t_A-T)& p(\widehat{t}_A-t_B-T)&p(\widehat{t}_A-t_A-2T)& p(\widehat{t}_A-t_B-2T)&\cdots \\[0.2cm]
p(\widehat{t}_B-t_A)& p(\widehat{t}_B-t_B)&p(\widehat{t}_B-t_A-T)& p(\widehat{t}_B-t_B-T)&p(\widehat{t}_B-t_A-2T)& p(\widehat{t}_B-t_B-2T)&\cdots \\[0.2cm]
p(\widehat{t}_A-t_A+T)& p(\widehat{t}_A-t_B+T)&p(\widehat{t}_A-t_A)& p(\widehat{t}_A-t_B)&p(\widehat{t}_A-t_A-T)& p(\widehat{t}_A-t_B-T)&\cdots \\[0.2cm]
p(\widehat{t}_B-t_A+T)& p(\widehat{t}_B-t_B+T)&p(\widehat{t}_B-t_A)& p(\widehat{t}_B-t_B)&p(\widehat{t}_B-t_A-T)& p(\widehat{t}_B-t_B-T)&\cdots \\[0.2cm]
p(\widehat{t}_A-t_A+2T)& p(\widehat{t}_A-t_B+2T)&p(\widehat{t}_A-t_A+T)& p(\widehat{t}_A-t_B+T)&p(\widehat{t}_A-t_A)& p(\widehat{t}_A-t_B)&\cdots \\[0.2cm]
p(\widehat{t}_B-t_A+2T)& p(\widehat{t}_B-t_B+2T)&p(\widehat{t}_B-t_A+T)& p(\widehat{t}_B-t_B+T)&p(\widehat{t}_B-t_A)& p(\widehat{t}_B-t_B)&\cdots \\[0.2cm]
\vdots              &\vdots              &\vdots              &\vdots              &\vdots              &\vdots              &\ddots
\end{smallmatrix}
\end{bmatrix}.
\end{eqnarray}

Moreover, we specify the following observations from \eqref{EquC4},
\begin{itemize}
\item[(i)] The coefficient matrix $\bm{\Gamma}$ is related to the real time offsets $t_A$ and $t_B$ as well as the estimated time offsets $\widehat{t}_A$ and $\widehat{t}_B$.
Thus, $\bm{\Gamma}$ is unknown to the receiver since $t_A$ and $t_B$ are unknown.
\item[(ii)] The noise $\bm{w}$ is colored, its covariance matrix is given by
\begin{eqnarray}\label{EquC10}
\bm{\Sigma(w)} =\sigma^2\bm{\Sigma}=\sigma^2
\begin{bmatrix}
\begin{smallmatrix}
1                   & p(\widehat{\tau})   & 0                     & p(\widehat{\tau}-T)   & \cdots & p(\widehat{\tau}-(N-1)T)   \\[0.2cm]
p(-\widehat{\tau})  & 1                   & p(-\widehat{\tau}-T)  & 0                     & \cdots & 0                          \\[0.2cm]
0                   & p(\widehat{\tau}+T) & 1                     & p(\widehat{\tau})     & \cdots & p(\widehat{\tau}-(N-2)T)    \\[0.2cm]
p(-\widehat{\tau}+T)& 0                   & p(-\widehat{\tau})    & 1                     & \cdots & 0                    \\[0.2cm]
\vdots              & \vdots              & \vdots                & \vdots                & \ddots & \vdots               \\[0.2cm]
p(-\widehat{\tau}+(N-1)T)  & 0                 & p(-\widehat{\tau}+(N-2)T)    & 0                 & \cdots & 1
\end{smallmatrix}
\end{bmatrix},
\end{eqnarray}
where $\widehat{\tau}=\widehat{t}_A-\widehat{t}_B$. Note that $\bm{\Sigma(w)}$ is only related to the estimated time offsets $\widehat{t}_A$ and $\widehat{t}_B$,
and hence is known to the receiver.
\item[(iii)] When the estimation is exact, i.e., $\widehat{t}_A=t_A$ and $\widehat{t}_B=t_B$, we have $\bm{\Gamma = \Sigma}$.
\end{itemize}

\subsection{Equivalent whitened model}
The discrete-time model in \eqref{EquC4} is essentially a time-varying ISI channel model, where the coefficients in the matrix $\bm{\Gamma}$ capture the severities of ISI: the more bib-zero off-diagonal elements the $\bm{\Gamma}$ has, the more severe is the ISI. On this basis, the conventional optimal decoders designed for ISI channel applies to \eqref{EquC4} naturally.

However, we notice that the noise $\bm{w}$ is colored. For the optimal decoder designed for ISI channel, colored noise generally complicates the optimal metrics and the decoding structure. Let us take the maximum likelihood sequence estimator (MLSE) \cite{viterbi} for instance. MLSE is a well-known optimal decoder for ISI channel subject to simple implementation by a Viterbi algorithm operating on a trellis. Given an input sequence with white noise, the metric in each transition reduces to a minimum Euclidean distance metric, and the decoding can be operated recursively since successive transitions are independent \cite{viterbi,ViterbiDetail}. Colored noise, on the other hand, would destroy this simple algorithm.

To retain the simplicity, we whiten the noise $\bm{w}$ before the samples are fed into optimal decoders.
Appendix B proves that the matrix $\bm{\Sigma}$ in \eqref{EquC10} is symmetric and positive definite. This suggests that the noise can be whitened using Cholesky decomposition as that in \eqref{EquB15}.

First, we factorize $\bm{\Sigma:\Sigma=U^T U}$, where $\bm{U}$ is a full-ranked upper triangular matrix. Then, we left multiply \eqref{EquC4} by $\bm{U^{-T}}$, giving,
\begin{eqnarray}\label{EquC11}
\bm{U^{-T}y}\!\!\!\!&=&\!\!\!\!\bm{U^{-T}\Gamma H x + U^{-T}w}\nonumber\\
&\overset{(a)}{=}&\!\!\!\!\bm{U^{-T}(\Sigma+\Delta)Hx+U^{-T}w}  ~~~~~~\overset{\textup{def}}{=}~~~~~~  \bm{\bar{y}=UHx+{\epsilon}+\bar{w}}~~~~~~~~~~~\\
&=&\!\!\!\!\bm{UHx+U^{-T}\Delta Hx+U^{-T}w},\nonumber
\end{eqnarray}
The structure of \eqref{EquC11} is explained and detailed below:
\begin{itemize}
\item[(i)] The transformed noise $\bm{\bar{w}}$ is now i.i.d. Gaussian noise with zero mean and covariance matrix $\sigma^2\bm{I}$ since $\mathbb{E}[\bm{\bar{w}}\bm{\bar{w}^H}]$ =$\mathbb{E}[\bm{U^{-T}ww^HU^{-1}}]=\sigma^2\bm{U^{-T}\Sigma U^{-1}}=\sigma^2\bm{I}$. Thus, \eqref{EquC11} is the whitened discrete-time model corresponding to \eqref{EquC4}.
\item[(ii)] In step (a), the matrix $\bm{\Gamma}$ is broken into the sum of two parts: $\bm{\Sigma}$ and $\bm{\Delta}$. Notice that if the estimation is exact (i.e. $\widehat{t}_A=t_A$ and $\widehat{t}_B=t_B$), we have $\bm{\Gamma = \Sigma}$. In other words, $\bm{\Delta}$ is an error matrix caused by inaccurate estimation.
\item[(iii)] Recall that the matrix $\bm{\Gamma}$ captures the severities of ISI. Compare matrix $\bm{U}$ with $\bm{\Gamma}$, it is interesting to note that the whitening transformation not only whitens the noise but also reduces complexity since the matrix $\bm{U}$ is an upper triangular matrix.
\end{itemize}

In \eqref{EquC11}, we further define an error vector $\bm{{\epsilon}=U^{-T}\Delta Hx}$. This error vector $\bm{{\epsilon}}$, caused by inaccurate estimation, is unknown to the receiver since the error matrix $\bm{\Delta}$ is unknown. Thus, the receiver will ignore $\bm{{\epsilon}}$ and assume
\begin{eqnarray}\label{EquC12}
\bm{\bar{y}~\approx ~ UHx +\bar{w}}.
\end{eqnarray}
This approximation is valid only when the estimation is accurate enough, or else, the decoding performance will suffer. As will be shown later, the estimation accuracy of our double baud-rate estimator enables a small $\bm{{\epsilon}}$ even with short preamble.
\subsection{Optimal decoder with truncated states}
An optimal PNC decoder aims to decode the most likely network-coded symbols in the presence of ISI. In general, two kinds of optimal decoders are possible for RRC-APNC \cite{PNCbook}:
\begin{itemize}
\item[(i)] The decoder that maximizes the joint APP of the sequence $\bm{s_{AB}=[s_A,s_B]^T}$, i.e. maximizing $\Pr(\bm{s_A,s_B}\mid \bm{\bar{y}})$. Specifically, the decoder first finds the most likely sequence $\widehat{\bm{s_{AB}}}$ given by \eqref{EquC13}, and then performs XOR operation to decode the network-coded sequence $\bm{\widehat{s}_{A\oplus B}}$.
    \begin{eqnarray}\label{EquC13}
    \widehat{\bm{s_{AB}}}=\arg\max_{\bm{s_A,s_B}}\Pr(\bm{x}\mid\bm{\bar{y}})
    \end{eqnarray}
\item[(ii)] The decoder that maximizes the marginal APP of each pair $\{s_A[n],s_B[n]\}$, i.e., maximizing $\Pr\{s_A[n],s_B[n]\mid \bm{\bar{y}}\}, n=1,2,...,N$. Specifically, the decoder first decodes the maximum APPs: $\Pr\{s_A[n],s_B[n]\mid \bm{\bar{y}}\}, n=1,2,...,N$, and then find the optimal $\widehat{s}_{A\oplus B}[n]$ given by
    \begin{eqnarray}\label{EquC14}
    \widehat{s}_{A\oplus B}[n]=\arg\max_{s_{A\oplus B}[n]} \sum_{\substack{s_A[n],s_B[n]:\\s_A[n]\oplus s_B[n]=s_{A\oplus B}[n]}} \Pr\Big\{ s_A[n],s_B[n] \mid \bm{\bar{y}}  \Big\}
    \end{eqnarray}
\end{itemize}
Decoders of type (i) can be realized by MLSE \cite{millerViterbi}, while decoders of type (ii) can be realized using BCJR algorithm \cite{DongnanBCJR} or SPA. For channel-coded PNC in this paper, we prefer decoders of type (ii) since it provides soft information that can be fed to the channel decoder. We further remark that BCJR algorithm and SPA are similar \cite{SPA}; BCJR is essentially a special application of SPA on trellis. Thus, this paper makes no distinctions between these two algorithms and focus on the SPA decoder.

\begin{figure}[ht]
  \centering
  \includegraphics[width=1\columnwidth]{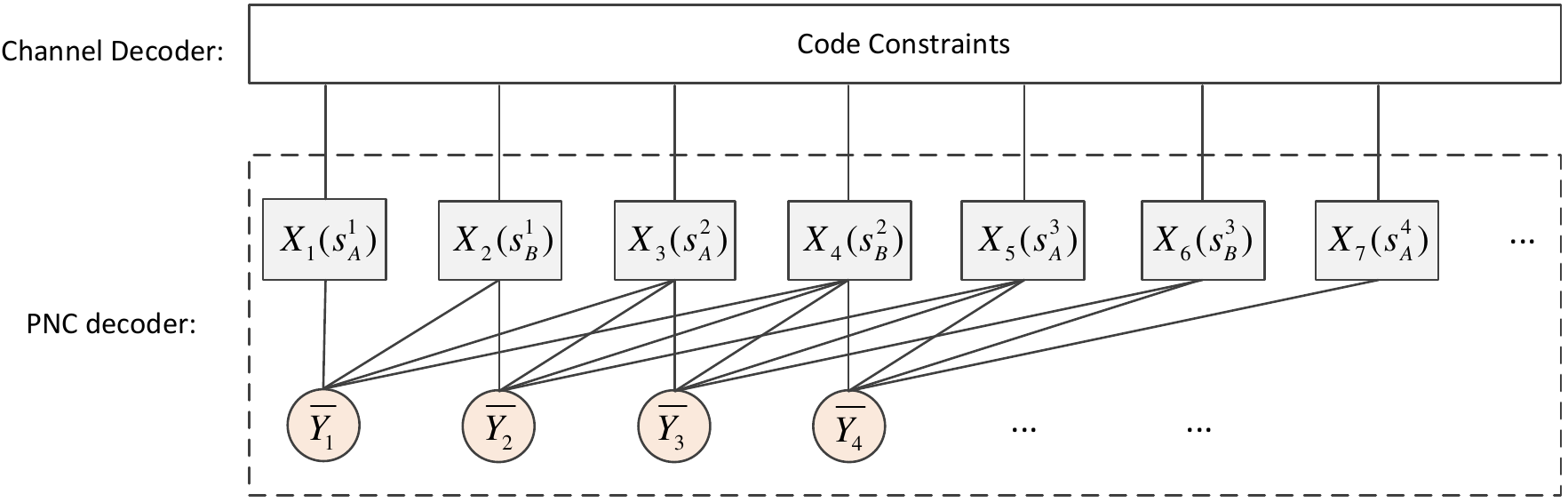}\\
  \caption{The graphical interpretation of \eqref{EquC12}, $X_i$ denotes the $i$-th variable node, $\overline{Y_i}$ denotes the $i$-th observation node. Each variable node $X_i$ represents a transmitted symbol $s_A[n]$ or $s_B[n]$ (abbreviate as $s_A^n$ or $s_B^n$). The state size in $\overline{Y_i}$ is truncated so that each $\overline{Y_i}$ is connected with $L$ $X_i$ ($L=4$ in this figure).}
\label{Fig5}
\end{figure}

The graphical interpretation of \eqref{EquC12} is given in Fig.\ref{Fig5}, where the error vector $\bm{{\epsilon}}$ is ignored, and the $i$-th entries in $\bm{x}$ and $\bm{\bar{y}}$ form the variable node $X_i$ and observation node $\overline{Y_i}$, respectively. In particular, we truncate the state size in each observation node to $q^L$ (i.e., the measure (ii) in the beginning of this section to enable feasible decoder), and hence each observation node only connects $L$ variable nodes. In other words, the decoder only considers the $L-1$ off-diagonals adjacent to the main diagonal beside the main diagonal in matrix $\bm{U}$, and the other off-diagonals are truncated. As a result, each variable node $X_i$ has $q$ states and each observation node $\overline{Y}_i$ has $q^L$ states.

\begin{figure}[ht]
  \centering
  \includegraphics[width=1\columnwidth]{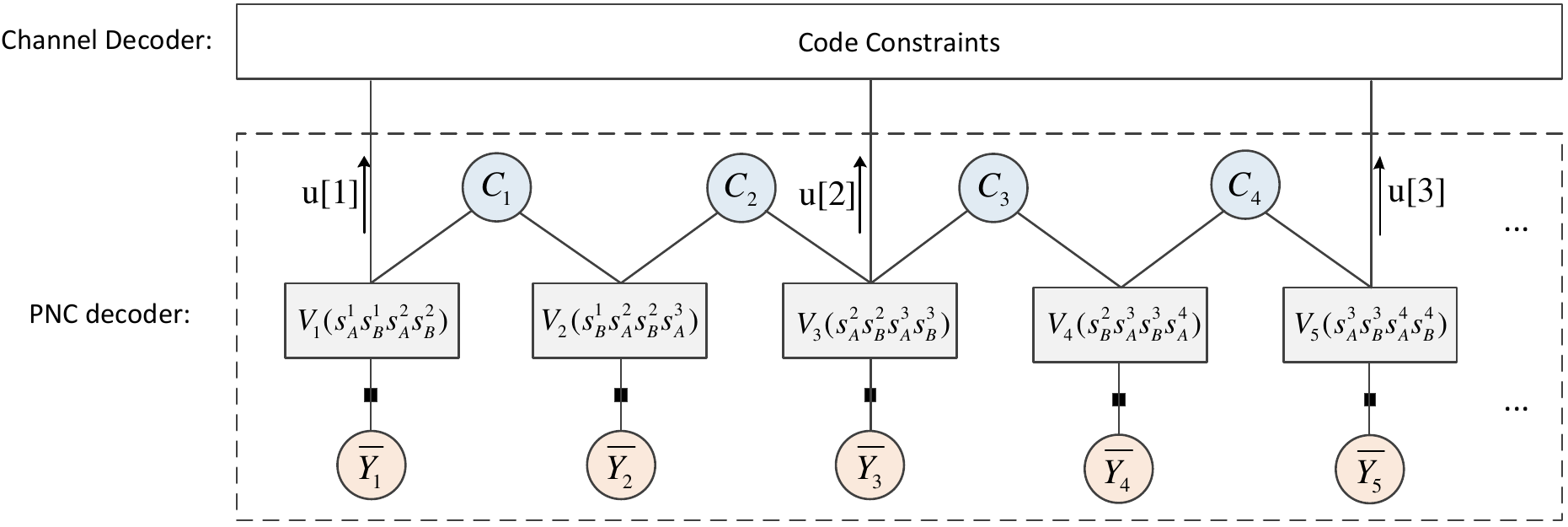}\\
  \caption{The equivalent acyclic factor graph corresponding to Fig.\ref{Fig5}, where $V_i$ denotes the $i$-th new variable node, $V_i=\{X_i,X_{i+1},X_{i+2},X_{i+3}\}$ ($L=4$); $C_i$ is the $i$-th check node; $\overline{Y_i}$ is the $i$-th observation node.}
\label{Fig6}
\end{figure}

The factor graph in Fig.\ref{Fig5} is cyclic. If we directly run SPA on this graph, iterative decoding is required and the resulting APP is not exact \cite{SPA}. However, we could first convert it into an acyclic graph.
A possible transformation is to cluster all the variable nodes that connect to the same observation node. Specifically, given Fig.\ref{Fig5}, we cluster $\{X_i,X_{i+1},X_{i+2},X_{i+3}\}$ and construct a new variable node $V_i=\{X_i,X_{i+1},X_{i+2},X_{i+3}\}$. The transformed factor graph after clustering is given in Fig.\ref{Fig6}, and the following variations can be observed:
\begin{itemize}
\item[(i)] The link between $V_i$ and $\overline{Y}_i$ is now singly-connected, and each new variable node $V_i$ has $q^L$ states.
\item[(ii)] Constraints among $V_i$ emerge, because common symbols are contained in adjacent $V_i$. These constraints are expressed through the check nodes $C_i$ in Fig.\ref{Fig6},\footnote{Note that the check on the adjacent two variable nodes is enough to depict all the constraints. Adding more check node would only result in redundancy and unnecessary cycles, hence degrade the performance. For instance, there is intersection between $V_1$ and $V_3$ but we do not need to check these two variable nodes since the constrains are embedded in check nodes $C_1$ and $C_2$ already.} whose corresponding check function $\psi_i$ is given by
    \begin{eqnarray}\label{EquC15}
    \psi_i=
    \begin{cases}
        1& \text{if the values of all the common symbols between $V_i$ and $V_{i+1}$ are equal}\\
        0& \text{if at least one common symbol between $V_i$ and $V_{i+1}$ differs in value}
    \end{cases}
    \end{eqnarray}
    For instance, the common symbols between $V_1$ and $V_2$ are $\{s_B^1,s_A^2,s_B^2\}$. Thus, $\psi_1=1$ for $V_1=\{b_1,1,1,1\},V_2=\{1,1,1,b_2\}$, and $\psi_1=0$ for $V_1=\{b_1,1,1,1\},V_2=\{-1,1,1,b_2\}$, where $b_1$ and $b_2$ can be either $1$ or $-1$.
\item[(iii)] This graph has a tree structure. This implies that we could start the computation from the leaf nodes (for this graph, the two boundary nodes on the left and right). Each message needs to be computed once and only once, after which the exact APP can be obtained \cite{SPA,PNCbook}.
\end{itemize}

Given the transformed acyclic graph in Fig.\ref{Fig6}, we can then run SPA to maximize the APPs $\Pr\{s_A[n],s_B[n]\mid\bm{\bar{y}}\}, n=1,2,...,N$.  The processing of SPA contains a forward (left-to-right) message passing and a backward (right-to-left) message passing. For the forward message passing, at each variable node, we take the product over all the incoming messages to form the outgoing message; at each check node, we sum up all the messages that link to the same state in the next variable node.

To simplify notations, we denote the $L$ variates in $V_i$ by $\{\omega_i^1, \omega_i^2, ..., \omega_i^L\}$, where each variate can be either $1$ or $-1$. Thus, each $V_i$ has $2^L$ states, and we define $V_i$ to be in state $(b_1,b_2,...,b_L)$ if $\omega_i^1=b_1$, $\omega_i^2=b_2$, ..., $\omega_i^L=b_L$, where $b_1, b_2, ..., b_L\in\{1,-1\}$. For instance in Fig.\ref{Fig6}, for $V_1$, we have $\omega_1^1=s_A^1$, $\omega_1^2=s_B^1$, $\omega_1^3=s_A^2$, $\omega_1^4=s_B^2$. Thus, $V_i$ is in state (1,1,1,1) when $\omega_1^1=\omega_1^2=\omega_1^3=\omega_1^4=1$.

{\bf Initialization}. First, we calculate the posterior probability of each state in $V_i$ based on the corresponding observation node $\overline{Y_i}$. That is, the probability that $V_i$ is in state $(b_1,b_2,...,b_L)$ is given by
\begin{eqnarray}\label{EquC16}
p_i(b_1,b_2,...,b_L)\!\!\!\!&=&\!\!\!\!\Pr(\omega_i^1=b_1,\omega_i^2=b_2,...,\omega_i^L=b_L\mid\bm{\bar{y}})\nonumber\\
&=&\!\!\!\!\kappa_i \exp\Bigg\{ -\frac{1}{2\sigma^2}\Big |\bar{y}[i]-\sum_{l=0}^{L-1}\bm{U}(i,i+l)\bm{H}(i+l,i+l)b_{l+1}  \Big |^2     \Bigg\}
\end{eqnarray}
where $\kappa_i$ is the normalization factor; $\bm{U}(i,j)$ or $\bm{H}(i,j)$ is the $(i,j)$-th entry of matrix $\bm{U}$ or $\bm{H}$ in \eqref{EquC12}.

\begin{figure}[ht]
  \centering
  \subfigure[Forward propagation]
  {
  \label{Fig7a}
  \includegraphics[width=0.35\columnwidth]{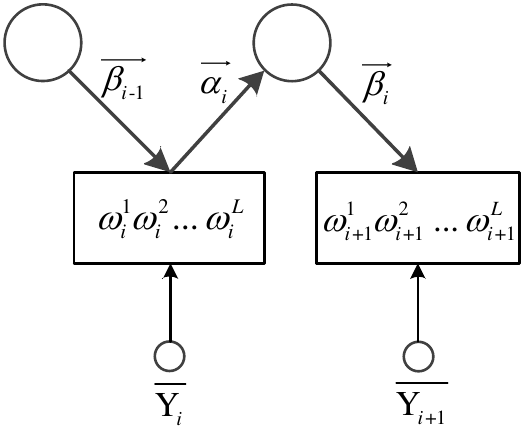}
  }
  \subfigure[Backward propagation]
  {
  \label{Fig7b}
  \includegraphics[width=0.35\columnwidth]{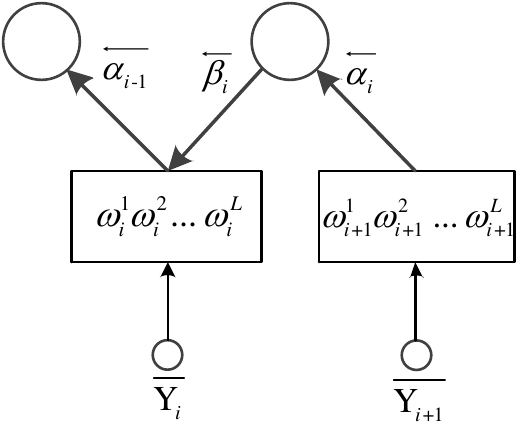}
  }
  \caption{The forward and backward propagations in SPA.}
  \label{Fig7}
\end{figure}

{\bf Forward propagation}. The ``sum'' and ``product'' operations in the forward propagation are shown in Fig.\ref{Fig7a}, where the messages relevant to state $(b_1,b_2,...,b_L)$ output from the $i$-th variable node and check node are defined as $\overrightarrow{\alpha_i}(b_1,b_2,...,b_L)$ and $\overrightarrow{\beta_i}(b_1,b_2,...,b_L)$, respectively.

Specifically, we have
\begin{eqnarray}
\label{EquC17}
\overrightarrow{\alpha_i}(b_1,b_2,...,b_L)\!\!\!\!&=&\!\!\!\!\overrightarrow{\beta_i}(b_1,b_2,...,b_L)p_i(b_1,b_2,...,b_L);\\
\label{EquC18}
\overrightarrow{\beta_i}(b_1,b_2,...,b_{L-1},\pm 1)\!\!\!\!&=&\!\!\!\!\kappa'\sum_{b_1}\overrightarrow{\alpha_i}(b_1,b_2,...,b_L),
\end{eqnarray}
where $\kappa'$ is a normalization factor, and $p_i(b_1,b_2,...,b_L)$ is given in \eqref{EquC16}.

{\bf Backward propagation}. The ``sum'' and ``product'' operations in the backward propagation are shown in Fig.\ref{Fig7b}, where the messages relevant to state $(b(b_1,b_2,...,b_L)$ output from the $i$-th variable node and check node are defined as $\overleftarrow{\alpha_i}(b_1,b_2,...,b_L)$ and $\overleftarrow{\beta_i}(b_1,b_2,...,b_L)$, respectively.

Specifically, we have
\begin{eqnarray}
\label{EquC19}
\overleftarrow{\beta_i}(\pm 1,b_2,...,b_L)\!\!\!\!&=&\!\!\!\!\kappa''\sum_{b_L}\overleftarrow{\alpha_i}(b_1,b_2,...,b_L);\\
\label{EquC20}
\overleftarrow{\alpha_{i-1}}(b_1,b_2,...,b_L)\!\!\!\!&=&\!\!\!\!\overleftarrow{\beta_i}(b_1,b_2,...,b_L)p_i(b_1,b_2,...,b_L),
\end{eqnarray}
where $\kappa''$ is a normalization factor, and $p_i(b_1,b_2,...,b_L)$ is given in \eqref{EquC16}.

{\bf Marginalization}. After one forward propagation from left to right and one backward propagation from right to left, the APPs of each state in variable nodes now converge, giving
\begin{eqnarray}\label{EquC21}
\ddot{p}_i(b_1,b_2,...,b_L)=\overrightarrow{\beta_{i-1}}(b_1,b_2,...,b_L)\overleftarrow{\beta_i}(b_1,b_2,...,b_L)p_i(b_1,b_2,...,b_L), ~~~i=1,2,...,2N.
\end{eqnarray}

Further, extracting the odd entries of \eqref{EquC21}, we have
\begin{eqnarray}\label{EquC22}
\Pr\{s_A[n]=b_1,s_B[n]=b_2\mid \bm{\bar{y}} \}=\sum_{b_3,b_4}\ddot{p}_{2n+1}(b_1,b_2,...,b_L), ~~~n=1,2,...,N.
\end{eqnarray}
Certainly, SPA can also operate in log-domain to avoid the numerical issues (i.e., the log-SPA \cite{LOGMAP}).

{\bf Decoding}. For a non-channel-coded system, we directly decode the most likely network-coded symbol $s_{A\oplus B}[n]$ from \eqref{EquA7}. For a channel-coded system, we could perform XOR-CD decoding or Jt-CNC decoding \cite{PNCbook}, and turbo decoding is also possible under the framework \cite{Turbo}. However, comparing the performance of different decoders is beyond the scope of this paper. For simplicity, we perform XOR-CD decoding based on \eqref{EquC22}. Specifically, instead of decoding $s_{A\oplus B}[n]$ directly, we calculate the soft information as
\begin{eqnarray}\label{EquC23}
u[n]\!\!\!\!&=&\!\!\!\!\frac{\Pr\{s_{A\oplus B}[n]=0\}}{\Pr\{s_{A\oplus B}[n]=1\}}\nonumber\\
&=&\!\!\!\!\frac{\sum_{b_1\oplus b_2=0}\Pr\{s_A[n]=b_1,s_B[n]=b_2\mid \bm{\bar{y}} \}}{\sum_{b_1\oplus b_2=1}\Pr\{s_A[n]=b_1,s_B[n]=b_2\mid \bm{\bar{y}} \}}.
\end{eqnarray}
The soft information is then passed to the channel decoder, and we decode the network-coded symbols after several iterations inside the channel decoder.

\subsection{Numerical results}
Given the symbol misalignment estimators presented in Section III and the optimal PNC decoders presented in Section IV, there are four possible estimation-and-decoding solutions for RRC-APNC as shown in Table.~\ref{Table1}.
\begin{table}[!hbp]
\centering
\caption{Four possible estimation-and-decoding solutions for RRC-APNC.}
\label{Table1}
\begin{tabular}{|p{5cm}<{\centering}|p{5cm}<{\centering}|p{5cm}<{\centering}|}
  \hline
   & Baud-rate estimation & Double baud-rate estimation \\
   \hline
  Baud-rate decoding & Solution I & Solution II \\
  \hline
  Double baud-rate decoding & Solution III & Solution IV \\
  \hline
\end{tabular}
\end{table}
Specifically, solution I corresponds to a baud-rate sampling system, where estimation and decoding are based on the baud-rate samples. Solution IV corresponds to a double baud-rate sampling system, where estimation and decoding are based on double baud-rate samples. Solution II and III are the simplification of solution IV, where we first sample the received signal at double baud-rate but drop half of the samples corresponding to the preamble part (solution III) or data part (solution II).

This subsection studies these four estimation-and-decoding solutions to evaluates the overall performance of RRC-APNC when incorporating symbol misalignment estimation.
We assume the transmitted symbols $\bm{s_A}$ and $\bm{s_B}$ are BPSK modulated,\footnote{The simulations focus on the effect of symbol-misalignment estimation on the overall decoding performance, hence BPSK modulation is suffice. Besides, BPSK enables simpler SPA decoder (i.e., the modulation order of BPSK is $q=2$, and hence the state size in each observation node is $2^L$).} and both nodes A and B utilize the same $(1024, 512)$ LDPC code \cite{LDPC1,LDPC2}. The relative time offset $\tau$ (i.e., symbol misalignment) is uniformly distributed in $[-T/2, T/2)$.

\subsubsection{Truncation issue}
As indicated in Section IV, truncation in state size is inevitable in the design of practical RRC-APNC decoders. For each observation node (in Fig.\ref{Fig5} or Fig.\ref{Fig8}), we truncate the state size from $2^{2N}$ to $2^L$. The choice of $L$ is a tradeoff between performance and complexity. That is, given a fixed roll-off factor of the RRC pulse $\beta$, the greater the $L$, the less the information loss, but the more complicated the decoder. Thus, the first thing we need to figure out is to find the proper $L$ under various $\beta$.
\begin{figure}[ht]
  \centering
  \includegraphics[width=0.6\columnwidth]{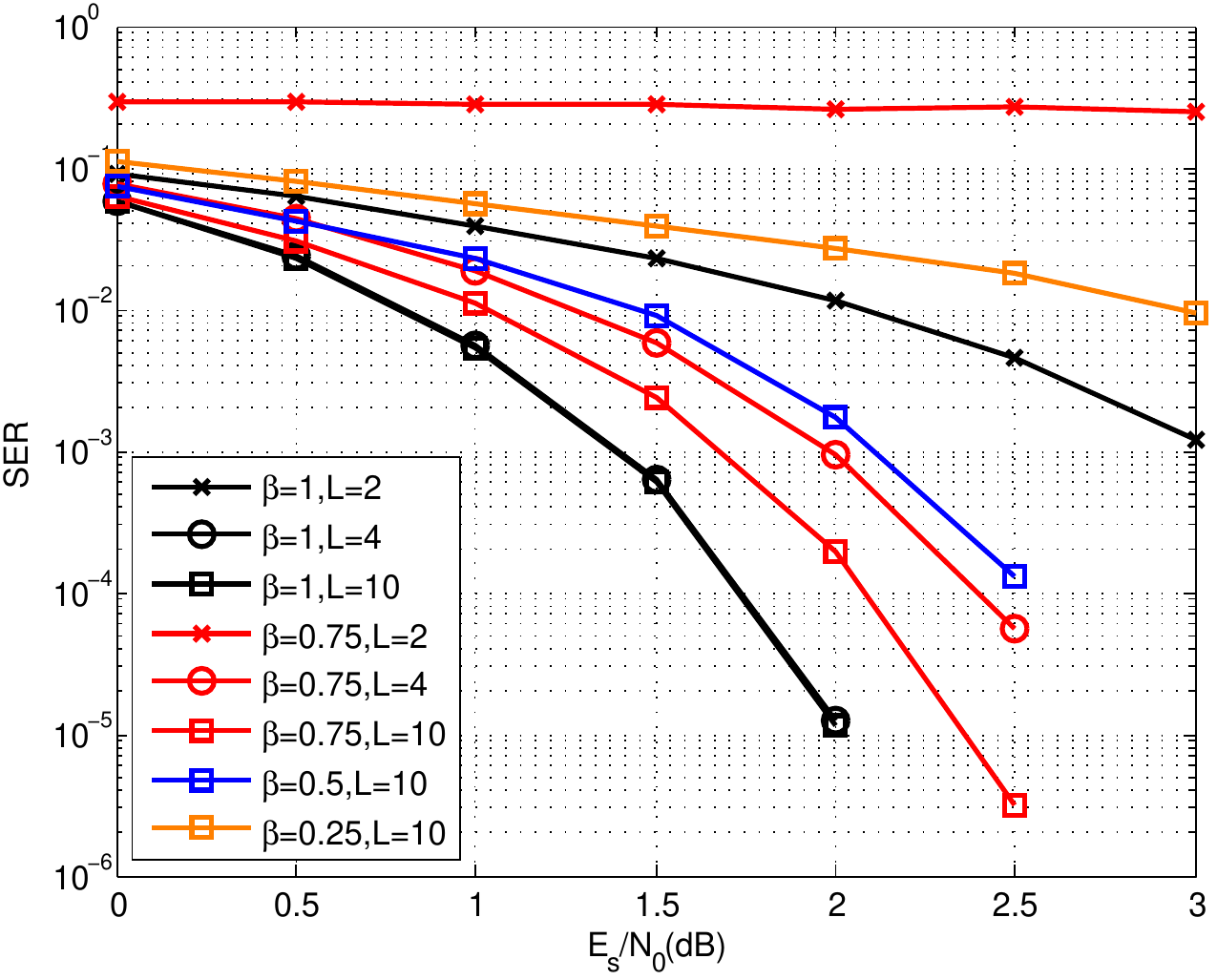}\\
  \caption{The SER of RRC-APNC with exact time offset estimation and double baud-rate decoding. We set $h_A=h_B=1$, $T=1$, the relative time offset $\tau$ is uniformly distributed in $[-0.5,0.5)$.}
\label{FigSim6}
\end{figure}

Assuming exact symbol-misalignment estimation, we investigate the effect of truncation (i.e., various $L$) on the symbol error rate (SER) of double baud-rate decoder.\footnote{We specify that the conclusions drawn in this simulation apply to the baud-rate decoder as well.} The simulation results are shown in Fig.\ref{FigSim6}. Intuitively, a smaller $\beta$ demands a greater $L$ to guarantee the decoding performance. Thus, we start from $\beta=1$ and increase $L$ from $2$ to $10$. As can be seen, the SER improves with the increase of $L$ when $L\leq 4$, whereas when $L>4$, no distinct improvement can be observed. This implies that $L=4$ is a good choice when $\beta=1$.
Then, fixing $\beta=0.75$, we increase $L$ from $2$ to $10$. The SER improves continuously with the increase of $L$. Further, if we increase $L$ beyond $10$, the SER no longer improves. Thus, we could set $L=10$ when $\beta=0.75$.

However, given $L=10$ and BPSK modulation, the state size in each observation node is $1024$. Such an enormous state size has already placed a heavy burden on the receiver, and hence further increasing $L$ may not be wise. Thus, for practical RRC-APNC, a relative large roll-off factor $\beta$ is preferred.

\subsubsection{Estimation-and-decoding}
\begin{figure}[t]
  \centering
  \includegraphics[width=0.8\columnwidth]{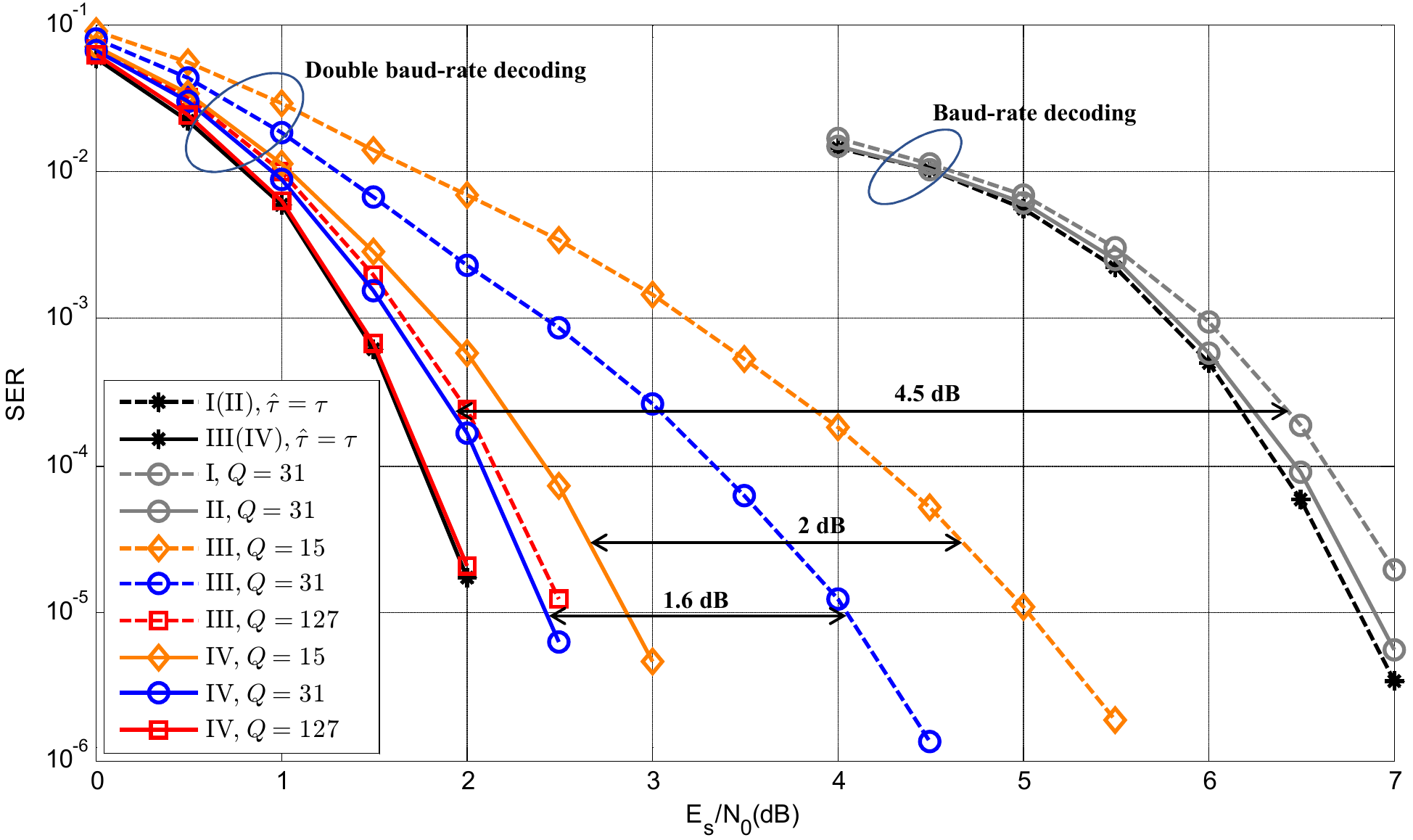}\\
  \caption{The SERs of the four estimation-and-decoding solutions in Table.\ref{Table1}. For estimators, we set $\beta=1$, $Q=15,31$ or $127$, $G=\lfloor Q/3\rfloor$, $d=4$; for decoders, we set $L=4$. Besides, $T=1$, $h_A=h_B=1$, the relative time offset $\tau$ is uniformly distributed in $[-0.5,0.5)$. The dashed curves mark the SERs of solutions using baud-rate estimator (i.e., solutions I and III) while the solid curves mark the SERs of solutions using double baud-rate estimator (i.e., solutions II and IV ).}
\label{FigSim7}
\end{figure}

Based on the observations in Fig.\ref{FigSim6}, we set $\beta=1$ and $L=4$ in this simulation and evaluate the performance of four estimation-and-decoding solutions for RRC-APNC in Table.~\ref{Table1}. The simulation results are shown in Fig.\ref{FigSim7}.

The first observation from Fig.\ref{FigSim7} is that, there is a gap between the SERs of solution I(II) (with baud-rate decoder) and solution III(IV) (with double baud-rate decoder). When $Q=31$, a double baud-rate sampling system (i.e., solution IV, $Q=31$) yields $4.5$ dB SER gains over a baud-rate sampling system (i.e., solution I, $Q=31$). Even with exact symbol-misalignment estimation (i.e., $\widehat{\tau}=\tau$), the $4.5$ dB gap between solutions I and IV retains (the two dark curves in Fig.\ref{FigSim7}).
This implies that, double baud-rate decoding is indispensable for RRC-APNC, especially when a large roll-off factor is used.

Then, we study the impact of different estimators on the decoding performance assuming the double baud-rate decoder is used in both cases (i.e., solutions III and IV).
Benchmarked against the exact misalignment-estimation case (solution III(IV), $\widehat{\tau}=\tau$), solution III suffers a $2.8$ dB SER loss when $Q=15$, and a $2$ dB loss when $Q=31$; whereas for solution IV, the SER loss is $0.8$ dB and $0.4$ dB when $Q=15$ and $31$, respectively. Further, when $Q=127$, solution IV almost achieves the benchmark, while solution III still suffers a $0.5$ dB loss.
\begin{figure}[t]
  \centering
  \includegraphics[width=0.6\columnwidth]{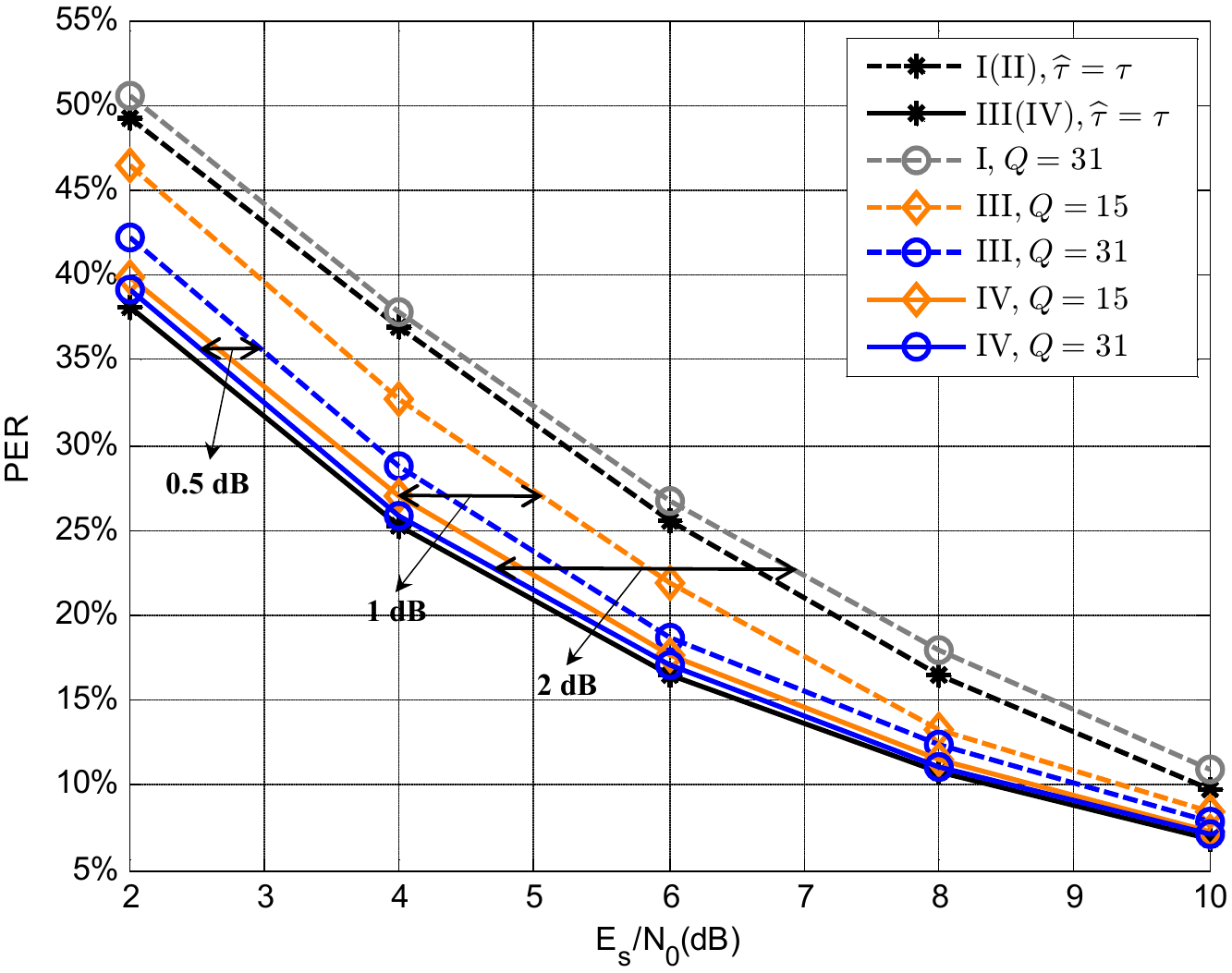}\\
  \caption{The PERs of the four estimation-and-decoding solutions in Table.\ref{Table1}. For estimators, we set $\beta=1$, $Q=15$ or $31$, $G=\lfloor Q/3\rfloor$, $d=4$; for decoders, we set $L=4$. Besides, $T=1$, the relative time offset $\tau$ is uniformly distributed in $[-0.5,0.5)$ and the channel gains $h_A$ and $h_B$ are Rayleigh distributed. The dashed curves mark the SERs of solutions using baud-rate estimator (i.e., solutions I and III) while the solid curves mark the SERs of solutions using double baud-rate estimator (i.e., solutions II and IV ).
  }
\label{FigSim8}
\end{figure}

Overall, solution IV yields considerable SER improvements over solition III under the same preamble length (e.g., $2$ dB when $Q=15$ and $1.6$ dB when $Q=31$). For the same SER performance, solution IV only requires a quarter of the overhead (i.e., preamble length) that solution III needs.

Fig.\ref{FigSim8} shows the decoding performances of various estimation-and-decoding solutions in a flat Rayleigh fading channel. As with Fig.\ref{FigSim4}, SER performance does not fully characterize the decoding performance in a Rayleigh fading channel, because the large SER of packets under deep fading may pull down the overall average performance. Thus, we analyze the packet error rate (PER) instead.

As can be seen, a double baud-rate sampling system (i.e., solution IV) yields $2$ dB PER gains over a baud-rate sampling system (i.e., solution I) when $Q=31$.
Solution IV yields $1$ dB PER gains over solution III when $Q=15$, and $0.5$ dB PER gains over solution III when $Q=31$.
Basically, in terms of PER performance, solution IV achieves the benchmark when $Q=31$ in a flat Rayleigh fading channel.

\section{Conclusion}
This paper investigated the joint problem of symbol-misalignment estimation and decoding in asynchronous physical-layer network coding systems using the root-raised-cosine pulse (RRC-APNC).
For symbol-misalignment estimation, we put forth an optimal double baud-rate estimator that is substantially more accurate than prior schemes under both AWGN and Rayleigh fading channels.
For example, the mean-square-error (MSE) gains of the double baud-rate estimator over the baud-rate estimator can be up to $8$ dB.
For PNC decoding, we devised optimal APNC decoders under inaccurate symbol-misalignment estimates. In particular, we proved and presented a new whitening transformation to whiten the noise of the double baud-rate samples. We investigated the decoding performance of various estimation-and-decoding solutions for RRC-APNC. Our results indicate that, in order to reduce decoder complexity while maintaining good performance, a large roll-off factor of the RRC pulse is preferable; and that with a large roll-off factor, double baud-rate decoding is indispensable. Overall, the scheme that makes use of double baud-rate samples in both symbol-misalignment estimation and PNC decoding has much better performance than other solutions for RRC-APNC.

\appendices
\section{Derivations on the Outcome of Cross-correlation}
\subsection{Cross-correlation results}
Given the double baud-rate samples $\bm{\bar{y}^d}$ and $\bm{z^d_A}$ in \eqref{EquB8} and \eqref{EquB9}, we derive the cross-correlation results, i.e., $\bm{c^d_A}$ in \eqref{EquB11}, in this subsection.

First, we denote by $\bm{\bar{y}_e}, \bm{\bar{y}_o}$ and
$\bm{z_e}, \bm{z_o}$ the even and odd entries of $\bm{\bar{y}^d}$ and $\bm{z^d_A}$, respectively. Thus, we have
\begin{eqnarray}
\label{AppA1}
\bar{y}_e[n]\!\!\!\!&=&\!\!\!\!h_A\sum_l s_A[n-l]p'(lT-t_A) + h_B\sum_l s_B[n-l]p'(lT-t_B)+ w_e[n],\\
\label{AppA2}
\bar{y}_o[n]\!\!\!\!&=&\!\!\!\!h_A\sum_l s_A[n-l]p'(lT+\frac{1}{2}T-t_A) + h_B\sum_l s_B[n-l]p'(lT+\frac{1}{2}T-t_B)+ w_o[n],\\
\label{AppA3}
z_e[n]\!\!\!\!&=&\!\!\!\!\sum_l z_A[n-l]p'(lT),\\
\label{AppA4}
z_o[n]\!\!\!\!&=&\!\!\!\!\sum_l z_A[n-l]p'(lT+\frac{1}{2}T),
\end{eqnarray}
where $\bm{w_e}, \bm{w_o}$ are the even and odd entries of the i.i.d. Gaussian noise $\bm{\bar{w}^d}$ in \eqref{EquB8}.

To obtain $\bm{c^d_A}$, we then cross-correlate $\bm{\bar{y}^d}$ with ${\bm{z^d_A}}^*$. Thus, the even and odd entries of $\bm{c^d_A}$, i.e., $\bm{c_e}$ and $\bm{c_o}$, are given by
\begin{eqnarray}
\label{AppA5}
c_e[m]\!\!&=&\!\!\sum_{n=0}^{Q-1}z^*_e[n]y_e[n\!+\!m]+\sum_{n=0}^{Q-1}z^*_o[n]y_o[n\!+\!m],\\
\label{AppA6}
c_o[m]\!\!&=&\!\!\sum_{n=0}^{Q-1}z^*_e[n]y_o[n\!+\!m]+\sum_{n=0}^{Q-1}z^*_o[n]y_e[n\!+\!m\!+\!1].
\end{eqnarray}

The first term on the R.H.S. of \eqref{AppA5} gives
\begin{eqnarray}\label{AppA7}
&&\!\!\!\!\!\!\!\!\!\!\!\!\sum_{n=0}^{Q-1}z^*_e[n]y_e[n\!+\!m]\nonumber\\
&\overset{(a)}{=}&\!\!\sum_{n=0}^{Q-1}\sum_lz^*_A[n\!-\!l]p'(lT)h_A\sum_{l'}z_A[n\!+\!m\!-\!l']p'(l'T\!-\!t_A) +\sum_{n=0}^{Q-1}z^*_e[n]w_e[n+m]\nonumber\\
&\overset{(b)}{=}&\!\!Qh_A\sum_l p'(lT) p'(lT+mT-t_A)+\widetilde{w}^1_e[m],
\end{eqnarray}
where (a) follows since we are only interested in the preamble parts embedded in $\bm{s_A}$ and $\bm{s_B}$; (b) follows by setting $l'=l+m$.

In the same way, the second term on the R.H.S. of \eqref{AppA5} can be simplified as
\begin{eqnarray}\label{AppA8}
&&\!\!\!\!\!\!\!\!\!\!\!\!\sum_{n=0}^{Q-1}z^*_o[n]y_o[n\!+\!m]=Qh_A\sum_l p'(lT\!+\!\frac{T}{2})p'(lT\!+\!\frac{T}{2}\!+\!mT\!-\!t_A) +\widetilde{w}^2_e[m].\nonumber
\end{eqnarray}

Substituting \eqref{AppA7} and \eqref{AppA8} into \eqref{AppA5}, we have
\begin{eqnarray}\label{AppA9}
\!\!\!\!c_e[m]\!\!\!\!\!&=&\!\!\!\!\!
Qh_A\sum_l
\Big[p'(lT)p'(lT+mT-t_A) +p'(lT\!+\!\frac{T}{2})p'(lT\!+\!\frac{T}{2}\!+\!mT\!-\!t_A)\Big]+\widetilde{w}_e[m],
\end{eqnarray}
where $\bm{\widetilde{w}_e=\widetilde{w}^1_e+\widetilde{w}^2_e}$.

Similarly, \eqref{AppA6} can be refined as
\begin{eqnarray}\label{AppA10}
\!\!\!\!c_o[m]\!\!\!\!\!&=&\!\!\!\!\!
Qh_A\sum_l
\Big[p'(lT)p'(lT\!+\!\frac{T}{2}\!+\!mT\!-\!t_A) +p'(lT\!+\!\frac{T}{2})p'(lT\!+\!T\!+\!mT\!-\!t_A)\Big]\!+\!\widetilde{w}_o[m].
\end{eqnarray}

Given \eqref{AppA9} and \eqref{AppA10}, we combine the even and odd entries of $\bm{c^d_A}$, yields
\begin{eqnarray}\label{AppA11}
c^d_A[m]\!\!\!\!\!&=&\!\!\!\!\!
Qh_A\sum_l\Big[p'(lT)p'(lT\!+\!\frac{m}{2}T\!-\!t_A) +p'(lT\!+\!\frac{T}{2})p'(lT\!+\!\frac{m\!+\!1}{2}T\!-\!t_A)\Big]+\widetilde{w}^d_A[m].
\end{eqnarray}

\subsection{The noise issue}
In this subsection, we study the noise term in \eqref{AppA11}. We will prove that $\bm{\widetilde{w}^d_A}$ is colored while its even and odd entries, i.e. $\bm{\widetilde{w}_e}$ and $\bm{\widetilde{w}_o}$, are white. The auto-covariance function of $\bm{\widetilde{w}^d_A}$ is also given in this subsection.

First, we focus on the even entries $\bm{\widetilde{w}_e}$. Based on \eqref{AppA7} and \eqref{AppA8}, we have
\begin{eqnarray}\label{AppA12}
\widetilde{w}_e[m]\!\!&=&\!\!\widetilde{w}^2_e[m]+\widetilde{w}^2_e[m]\nonumber\\
&=&\!\!\sum_{n=0}^{Q-1}z^*_e[n]w_e[n+m]+\sum_{n=0}^{Q-1}z^*_o[n]w_o[n+m].
\end{eqnarray}

Thus, the auto-covariance function of $\bm{\widetilde{w}_e}$ is given by
\begin{eqnarray}\label{AppA13}
&&\!\!\!\!\!\!\!\!\!\!\!\!\mathbb{E}\big[\widetilde{w}_e[m]\widetilde{w}_e[m+j]\big]\\
&=&\!\!\sum_{n=0}^{Q-1}z_e[n]w^*_e[n+m]\sum_{n'=0}^{Q-1}z^*_e[n']w_e[n'\!+\!m\!+\!j] +\sum_{n=0}^{Q-1}z_o[n]w^*_o[n\!+\!m]\sum_{n'=0}^{Q-1}z^*_o[n']w_o[n'\!+\!m\!+\!j]\nonumber\\
&\overset{(a)}{=}&\!\!
2\sigma^2\sum_{n=0}^{Q-1}z_e[n]z^*_e[n-j]+2\sigma^2\sum_{n=0}^{Q-1}z_o[n]z^*_o[n-j],\nonumber
\end{eqnarray}
where in step (a), we set $n'=n-j$ and assume $N\gg j$.

Substituting \eqref{AppA3} and \eqref{AppA4} into \eqref{AppA13}, we have
\begin{eqnarray}\label{AppA14}
&&\!\!\!\!\!\!\!\!\!\!\!\!\mathbb{E}\big[\widetilde{w}_e[m]\widetilde{w}_e[m+j]\big]\nonumber\\
&=&\!\!2Q\sigma^2\sum_l\Big[~p'(lT)p'(lT-jT)+p'(lT\!+\!\frac{T}{2})p'(lT\!+\!\frac{T}{2}\!-\!jT)~\Big]\nonumber\\
&=&\!\!
2Q\sigma^2\sum_k p'(\frac{k}{2}T)p'(\frac{k}{2}T-jT),
\end{eqnarray}

To prove $\bm{\widetilde{w}_e}$ is white noise, we need to demonstrate that $\sum_k p'(\frac{k}{2}T)p'(\frac{k}{2}T-jT)=\delta(jT)$. The proof is given below.

\begin{proof}
According to Nyquist-Shannon sampling theorem, if the sampling rate is at or above the Nyquist rate, the continuous-time signal can be reconstructed from the discrete sampling points using the SINC function. Thus, if we sample the RRC pulse at double baud-rate, the original continuous RRC pulse could be reconstructed as
\begin{eqnarray}\label{AppA15}
p'(t)\!\!&=&\!\!\sum_kp'(\frac{k}{2}T)\delta(t-\frac{k}{2}T)\ast p_s(2t)\nonumber\\
&=&\!\!\sum_k p'(\frac{k}{2}T)p_s(2t-kT).
\end{eqnarray}

Since the convolution of two RRC pulses gives us an RC pulse, we have
\begin{eqnarray}\label{AppA16}
p(t)\!\!&=&\!\!p'(t)\ast p'(t)=\int_{-\infty}^{\infty}p'(\varphi)p'(t-\varphi)d\varphi\nonumber\\
&=&\!\!\sum_k\sum_{k'}p'(\frac{k}{2}T)p'(\frac{k'}{2}T)\!\int\! p_s(2\varphi\!-\!kT)p_s(2t\!-\!2\varphi\!-\!k'T)d\varphi\nonumber\\
&=&\!\!\sum_k\sum_{k'}p'(\frac{k}{2}T)p'(\frac{k'}{2}T)p_s(2t-kT-k'T).
\end{eqnarray}

Moreover, the RC pulse and SINC pulse satisfy the Nyquist ISI-free criterion, namely $p(jT)=\delta(jT),p_s(jT)=\delta(jT)$, where $j$ is an integer. Substituting this criterion into \eqref{AppA16}, we have
\begin{eqnarray}\label{AppA17}
\delta(jT)\!\!&=&\!\!p(jT)\nonumber\\
&=&\!\!\sum_k\sum_{k'}p'(\frac{k}{2}T)p'(\frac{k'}{2}T)p_s(2jT-kT-k'T)\nonumber\\
&\overset{(a)}{=}&\!\!\sum_k p'(\frac{k}{2}T)p'(\frac{k}{2}T-jT),
\end{eqnarray}
where (a) follows by setting $k'=2j-k$.
Eq.\eqref{AppA17} concludes this proof.
\end{proof}

Given \eqref{AppA17}, the the auto-covariance function of $\bm{\widetilde{w}_e}$ in \eqref{AppA14} is given by
\begin{eqnarray}\label{AppA18}
\mathbb{E}\big[\widetilde{w}_e[m]\widetilde{w}_e[m+j]\big]=2Q\sigma^2\delta(jT).
\end{eqnarray}

Similarly, we could further obtain
\begin{eqnarray}
\label{AppA19}
\mathbb{E}\big[\widetilde{w}_o[m]\widetilde{w}_o[m+j]\big]\!\!&=&\!\!2Q\sigma^2\delta(jT),\\
\label{AppA20}
\mathbb{E}\big[\widetilde{w}_e[m]\widetilde{w}_o[m+j]\big]\!\!&=&\!\!2Q\sigma^2p(jT+\frac{T}{2}).
\end{eqnarray}

From \eqref{AppA18}, \eqref{AppA19} and \eqref{AppA20}, we conclude that $\bm{\widetilde{w}_e}$ and $\bm{\widetilde{w}_o}$ are white, while $\bm{\widetilde{w}^d_A}$ is colored and its  auto-covariance function is given by
\begin{eqnarray}\label{AppA21}
\mathbb{E}\big[{\widetilde{w}^d_A[m]}^*\widetilde{w}^d_A[m+j]\big]=2Q\sigma^2p(\frac{j}{2}T).
\end{eqnarray}
\section{Proof on Positive Definite Matrix}
In this appendix, we prove that the matrix $\bm{\Sigma}$ in \eqref{EquC10} is positive definite. In particular, the matrix $\bm{\Sigma_0}$ in \eqref{EquB13} is a special case of $\bm{\Sigma}$ when $\widehat{\tau}=T/2$. Thus, $\bm{\Sigma_0}$ is also positive definite.
\begin{proof}
A matrix $\bm{A}$ is said to be positive definite if the scalar $\bm{b^T\!\!Ab}$ is positive for every non-zero column vector $\bm{b}$. Thus, we will prove that
$\forall~\bm{b}\!=\!\big[b_0,b_1,...,b_{2N-1}\big]^T\!\neq\!\bm{0},~\bm{b^T\Sigma b}>0$.

First, we have
\begin{eqnarray}\label{AppB1}
\bm{b^T\Sigma b}\!\!&=&\!\!\big[b_0,b_1,...,b_{2N-1}\big]\bm{\Sigma}\big[b_0,b_1,...,b_{2N-1}\big]^T\nonumber\\
&=&\!\!\sum_{i=0}^{2N-1}b^2_i+2\sum_{i=0}^{N-1}b_{2i-1}\sum_{j=0}^{N-1}b_{2i}p(\widehat{\tau}+jT-iT)
\end{eqnarray}

Then, we define a new function $v(t)$ as
\begin{eqnarray}\label{AppB2}
v(t)=\sum_{i=0}^{N-1}s_A[i]p'(t-jT)+\sum_{i=0}^{N-1}s_B[i]p'(t-jT-\widehat{\tau}).
\end{eqnarray}
In fact, $v(t)$ is the received noise-free signal at the relay and the energy of $v(t)$ is given by
\begin{eqnarray}\label{AppB3}
&&\!\!\!\!\!\!\!\!\!\!\!\!\Psi[v(t)]=\int_{-\infty}^{\infty}v^2(t)dt\nonumber\\
=&&\!\!\!\!\!\int_{-\infty}^{\infty}\Big[\sum_{i=0}^{N-1}s_A[i]p'(t\!-\!jT)+\sum_{i=0}^{N-1}s_B[i]p'(t\!-\!jT\!-\!\widehat{\tau})\Big]^2dt\nonumber\\
=&&\!\!\!\!\!\sum_{i=0}^{N-1}s^2_A[i]\int_{-\infty}^{\infty} p'^2(t\!-\!iT)dt+\sum_{i=0}^{N-1}s^2_B[i]\int_{-\infty}^{\infty} p'^2(t\!-\!iT\!-\!\widehat{\tau})dt\nonumber\\
&&\qquad\qquad\qquad\qquad+2\sum_{i=0}^{N-1}\sum_{j=0}^{N-1}s_A[i]s_B[j]\int_{-\infty}^{\infty} p'(t\!-\!iT)p'(t\!-\!jT\!-\!\widehat{\tau})dt\nonumber\\
\overset{(a)}{=}&&\!\!\!\!\!\sum_{i=0}^{N-1}s^2_A[i]+\sum_{i=0}^{N-1}s^2_B[i] +2\sum_{i=0}^{N-1}\sum_{j=0}^{N-1}s_A[i]s_B[j]p(jT\!-\!iT\!+\!\widehat{\tau}),
\end{eqnarray}
where (a) follows since $\int_{-\infty}^{\infty}p'^2(t-iT)dt=1$.

A key observation is that, if we take $\bm{s_A}$ and $\bm{s_B}$ as the even and odd entries of $\bm{b}$, \eqref{AppB3} is equivalent to \eqref{AppB1}. This implies that, given $\bm{b\neq 0}$, $\bm{b^T\Sigma b}$ equals to the energy of $v(t)$, i.e., $\Psi[v(t)]>0$. Thus, $\bm{\Sigma}$ is a positive definite matrix.
\end{proof}

\section{Optimal RRC-APNC Decoder Under Baud-rate Sampling}
This appendix presents the optimal RRC-APNC decoder under baud-rate sampling. In general, the design follows the steps in Section IV: (1) Construct the discrete-time model using the baud-rate samples at the symbol boundaries. Note that baud-rate sampling enables samples with white noise, hence no noise whitening is needed. (2) Construct the factor graph based on the discrete-time model and transform it into an acyclic graph. (3) Run SPA to maximize the APPs and decode.

First, we resample $y_R(t)$ to obtain the samples at the symbol boundaries of $\bm{s_A}$.\footnote{We can also obtain the samples at the symbol boundaries of $\bm{s_B}$. The decoder can be designed in the same way.} Giving,
\begin{eqnarray}\label{AppC1}
y^b_0[n]\!\!\!\!&=&\!\!\!\!{y}_R(t=nT+\widehat{t}_A)\nonumber\\
&=&\!\!\!\!h_A\sum_l s_A[n-l]p(lT+\widehat{t}_A-t_A)+h_B\sum_l s_B[n-l]p(lT+\widehat{t}_A-t_B)+w^b_0[n]\nonumber\\
&=&\!\!\!\!h_A s_A[n] +h_B\sum_l s_B[n-l]p(lT+\widehat{\tau})+\epsilon^b[n]+w^b_0[n],
\end{eqnarray}
where $\bm{w^b_0}$ is i.i.d. Gaussian noise, $w^b_0[n]\sim\emph{CN}(0,\sigma^2)$; $\bm{\epsilon^b}$ is an error vector as the $\bm{\epsilon}$ in \eqref{EquC11}.

The error vector $\bm{\epsilon^b}$, caused by inaccurate symbol-misalignment estimation, is a function of $t_A$ and $t_B$, and hence is unknown to the receiver. Thus, the receiver will ignore $\bm{\epsilon^b}$ and assume
\begin{eqnarray}\label{AppC2}
y^b_0[n]=h_A s_A[n] +h_B\sum_l s_B[n-l]p(lT+\widehat{\tau})+w^b_0[n].
\end{eqnarray}
We emphasize that this approximation is valid only when the symbol-misalignment estimation is accurate enough, or else, the decoding performance will suffer.

\begin{figure}[ht]
  \centering
  \includegraphics[width=1\columnwidth]{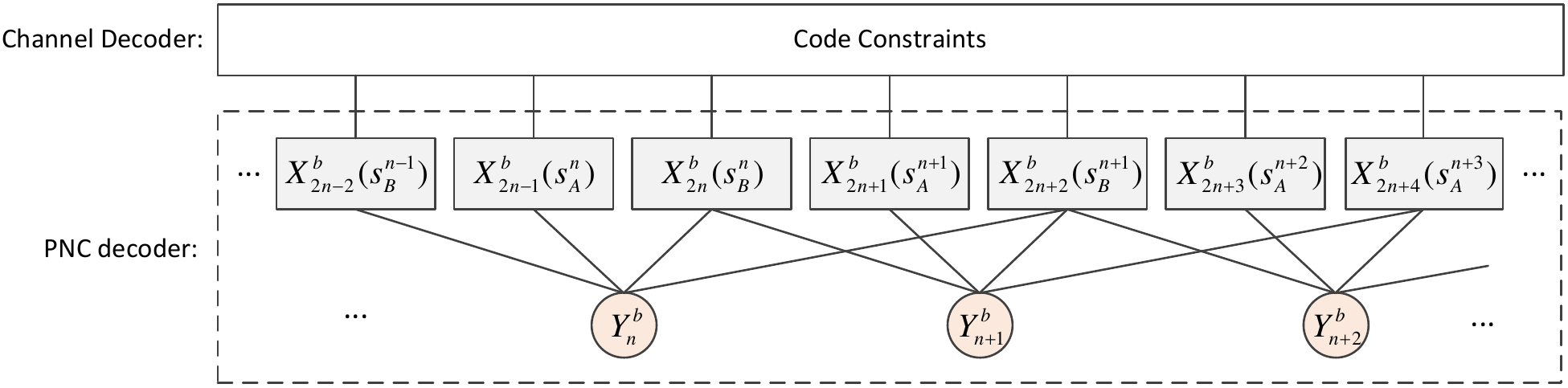}\\
  \caption{The graphical interpretation of \eqref{AppC2}, $X^b_i$ denotes the $i$-th variable node, $i=1,2,...,2N$; $Y^b_n$ denotes the $n$-th observation node, $n=1,2,...,N$. Each variable node $X^b_i$ represents a transmitted symbol $s_A[n]$ or $s_B[n]$ (abbreviate as $s_A^n$ or $s_B^n$). The state size in ${Y^b_n}$ is truncated so that each ${Y^b_n}$ is connected with $L$ $X^b_i$ ($L=4$ in this figure).}
\label{Fig8}
\end{figure}

Then, we interpret \eqref{AppC2} by a cyclic factor graph in Fig.\ref{Fig8}. In Fig.\ref{Fig8}, $y^b_0[n]$ forms the $n$-th observation node $Y^b_n$, $s_A[n]$ or $s_B[n]$ forms the odd-numbered observation node $X^b_{2n-1}, n=1,2,...,N$ or even-numbered observation node $X^b_{2n}, n=1,2,...,N$. The state size in each observation node is truncated so that each $Y^b_n$ only connects $1$ odd-numbered variable node and $L-1$ even-numbered variable nodes. For instance in Fig.\ref{Fig8}, $L=4$, and hence $Y^b_n$ connects
$\{X^b_{2n-2},X^b_{2n-1},X^b_{2n},X^b_{2n+2}\}$.

\begin{figure}[ht]
  \centering
  \includegraphics[width=1\columnwidth]{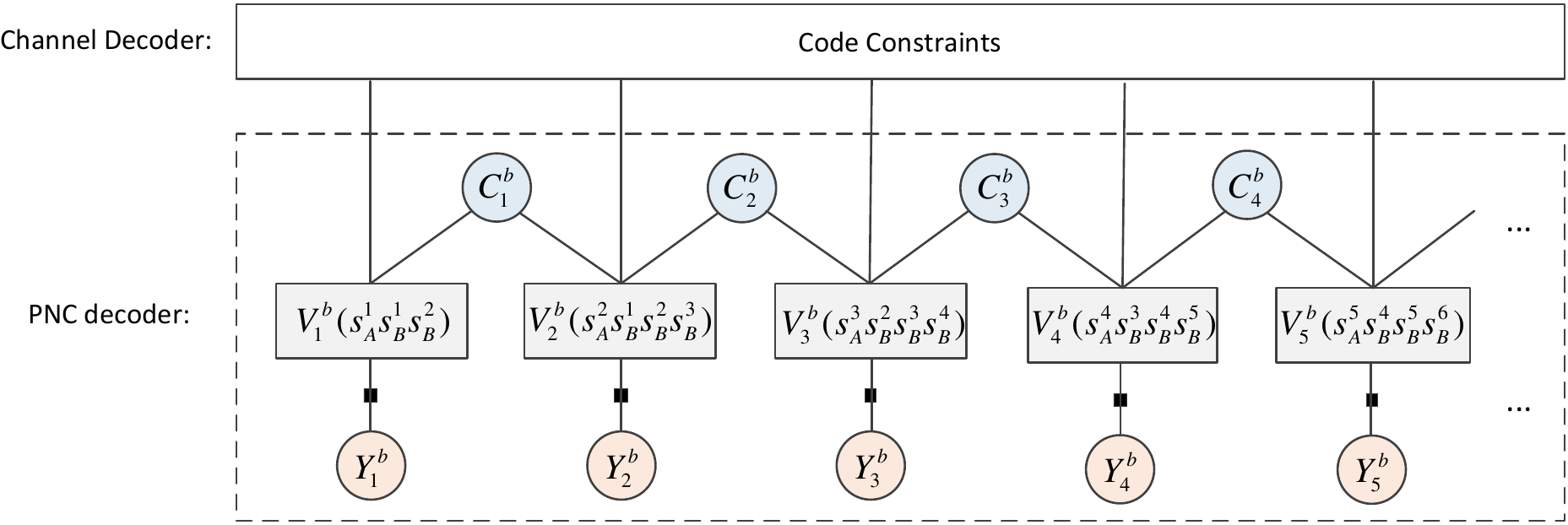}\\
  \caption{The equivalent acyclic factor graph corresponding to Fig.\ref{Fig8}, where $V^b_n$ denotes the $n$-th new variable node, $V^b_n=\{X^b_{2n-2},X^b_{2n-1},X^b_{2n},X^b_{2n+2}\}$ ($L=4$); $C^b_n$ is the $n$-th check node; ${Y^b_n}$ is the $n$-th observation node.}
\label{Fig9}
\end{figure}

Following the clustering in Fig.\ref{Fig6}, we group all the variable nodes that connect to the same observation node. Specifically, in Fig.\ref{Fig8}, we cluster $\{X^b_{2n-2},X^b_{2n-1},X^b_{2n},X^b_{2n+2}\}$ and construct a new variable node $V^b_n=\{X^b_{2n-2},X^b_{2n-1},X^b_{2n},X^b_{2n+2}\}$. The transformed acyclic factor graph is shown in Fig.\ref{Fig9}.

Given the transformed acyclic graph in Fig.\ref{Fig9}, we can then run SPA to maximize the APPs $\Pr\{s_A[n],s_B[n]\mid\bm{\bar{y}}\}, n=1,2,...,N$. The operations afterwards are similar to that in Section IV.

\ifCLASSOPTIONcompsoc
  \section*{Acknowledgments}
\else
  \section*{Acknowledgment}
\fi

The work of Y. Shao and S. C. Liew was supported by the General Research Funds (Project No. 14204714) established under the University Grant Committee of the Hong Kong Special Administrative Region, China.
The work of L. Lu was partially supported by the NSFC (Project No. 61501390) and partially supported by AoE grant E-02/08, established under the University Grant Committee of the Hong Kong Special Administrative Region, China.

\bibliographystyle{IEEEtran}
\bibliography{RRC_APNC}

\end{document}